\pdfoutput=1
\documentclass[a4paper,11pt]{article}
\usepackage{amsmath,amssymb,amsfonts}
\usepackage[multiple,stable]{footmisc}
\usepackage[amsmath,amsthm,thmmarks]{ntheorem}
\allowdisplaybreaks[4]
\usepackage[noconfig]{refstyle}
\usepackage[dvipsnames]{xcolor}
\usepackage[hyperfootnotes=false, linktocpage=true, colorlinks, citecolor=blue, linkcolor=blue, urlcolor=Maroon]{hyperref}
\usepackage{array,tabularx,booktabs,geometry,subfigure,graphicx,longtable}
\usepackage[bf]{caption}
\usepackage{tikz}
\usetikzlibrary{shapes,arrows}
\tikzstyle{block} = [rectangle, draw, text width=7em, text centered, rounded corners, minimum height=3em]
\usepackage{graphicx,color}
\usepackage{cite}

\let\eqref=\relax
\newref{f}{name={footnote~},Name={Footnote~},names={footnotes~},Names={Footnotes~}}
\newref{app}{name={appendix~},Name={Appendix~},names={appendixes~},Names={Appendixes~}}
\newref{tab}{name={table~},Name={Table~},names={tables~},Names={Tables~}}
\newref{ch}{name={chapter~},Name={Chapter~},names={chapters~},Names={Chapters~}}
\newref{sec}{name={section~},Name={Section~},names={sections~},Names={Sections~}}
\newref{fig}{name={figure~},Name={Figure~},names={figures~},Names={Figures~}}
\numberwithin{equation}{section}
\newcommand{\eref}[1]{(\ref{#1})}

\newcommand{\eeq}{\end{equation}}
\newcommand{\beq}{\begin{equation}}
\newcommand{\ba}{\begin{array}}
\newcommand{\ea}{\end{array}}

\newcommand{\be}{\begin{equation}}
\newcommand{\ee}{\end{equation}}
\newcommand{\bea}{\begin{equation}\begin{aligned}}	
\newcommand{\eea}{\end{aligned}\end{equation}}		

\newcommand{\tr}{\mathrm{tr}}

\newcommand{\iddots}{\mathinner{\mkern2mu\raise1pt\hbox{.}\mkern2mu \raise4pt\hbox{.}\mkern2mu\raise7pt\hbox{.}\mkern1mu}}

\providecommand{\id}{\leavevmode\hbox{\small$\mathrm{1}$\kern-3.8pt\normalsize$\mathrm{1}$}}
\def\fnote#1#2{\begingroup\def\thefootnote{#1}\footnote{#2}
     \addtocounter{footnote}{-1}\endgroup}

\begin{document}

\vspace{1cm}

\title{
       \vskip 40pt
       {\Large \bf Mapping moduli across heterotic conifolds}}
\author{Lara B. Anderson, James Gray, Sunit A. Patil, and Caoimh\'in Scanlon}
\date{}
\maketitle
\vspace{-0.8cm}
\begin{center} { {\it Physics Department, Robeson Hall, Virginia Tech,\\ Blacksburg, VA 24061, U.S.A.}}\\
\fnote{}{lara.anderson@vt.edu,jamesgray@vt.edu,sunitpatil@vt.edu,cscanlon@vt.edu}
\end{center}

\begin{abstract}
\noindent In this work, we provide evidence for a duality between $4$-dimensional Calabi-Yau compactifications of the heterotic string, in which the base manifolds are linked by a conifold transition. In recent work, a geometric proposal was put forward for how 5-branes and gauge bundles are carried across such transitions. It was observed that compactifications connected in this way lead to $4$-dimensional effective theories with the same massless spectrum. Here we provide much stronger evidence that these heterotic conifold transitions do indeed lead to dual theories. We construct a duality map between the field spaces of the two compactifications and use it to demonstrate the agreement of large numbers of holomorphic functions appearing in the definition of the effective theories. In an example, we show that 147,440 independent superpotential Yukawa couplings agree across the duality as holomorphic functions of the moduli. In certain special cases, the putative duality studied here reduces to the target space duality of $(0,2)$ gauged linear sigma models. 
\end{abstract}

\thispagestyle{empty}
\setcounter{page}{0}
\newpage

\tableofcontents

\section{Introduction}\label{sec:Intro}
Geometric transitions between Calabi-Yau manifolds \cite{Candelas:1988di,Green:1988wa,Green:1988bp,Candelas:1989js,Candelas:1989ug,Aspinwall:1993yb,Aspinwall:1993nu}, and string dualities associated to them, have proven to be remarkably powerful tools in exploring the string landscape and controlling string effective theories. In type IIB string theory, for example, flops between Calabi-Yau compactifications have changed our understanding of the moduli space of such vacua \cite{Aspinwall:1993nu}, and conifold singularities are one of the few limits in field space, away from asymptotic regions, where the effective theory takes a relatively simple form \cite{Strominger:1995cz,Greene:1995hu}. In the literature, tools such as these are often limited, in 4-dimensions, to theories with ${\cal N}=2$ supersymmetry or higher. 

Recent research has established an intriguing correspondence between $4$-dimensional ${\cal N}=1$ theories that are connected via conifold transitions between Calabi-Yau manifolds  \cite{Anderson:2022bpo}. This work described, geometrically, how bundles and branes in such compactifications can traverse the transitions (see \cite{Candelas:2007ac,Collins:2021azm,Collins:2021qqo} for some related work). In the context of heterotic string theory, compactifications that are linked in this way were shown to lead to theories which have the same massless spectrum of states. In certain special cases, these geometric transitions connect two theories related by the so called $(0,2)$ target space duality of gauged linear sigma models (GLSMs) \cite{Distler:1995bc,Blumenhagen:1997vt,Blumenhagen:1997cn,Blumenhagen:2011sq,Rahn:2011jw,Anderson:2016byt,Anderson:2019axt}. In \cite{Anderson:2022bpo} the important question was raised as to whether, more generally, these pairs of heterotic backgrounds are actually dual? In this work, our goal is to explore this possibility by attempting to define a mapping of 4-dimensional fields. This mapping is based upon constraints arising from the underlying geometry of the manifolds, branes, and bundles and the connection between them provided by the singular conifold limit. Using this mapping of fields we can verify matching of quantities in the effective theories beyond the spectrum. For any new purported duality of 4-dimensional ${\cal N}=1$ string compactifications, this is inherently difficult to explore due to the fact that many aspects of effective theories are not explicitly known (for example, the matter field K\"ahler potential \cite{McOrist:2016cfl,Candelas:2016usb,Candelas:2018lib,McOrist:2019mxh,Blesneag:2018ygh,Butbaia:2024tje,Constantin:2024yxh,Berglund:2024uqv,Gray:2025dao}). However, we show that the mapping we present  does indeed lead to highly non-trivial agreement between parts of the effective theory that can be computed on both sides of the duality.

\vspace{0.2cm}

The correspondence studied in \cite{Anderson:2022bpo} can be understood most directly by observing that there exist special curves from the point of view of a conifold pair of Calabi-Yau 3-folds, $X_D$ and $X_R$, which allow 5-branes wrapped on them to move through the conifold transition in an anomaly-consistent manner. We will refer to these curves as \emph{bridging curves} (and the branes that wrap them as \emph{bridging branes}). In more detail, the heterotic anomaly cancelation condition is as follows.
\begin{equation}
\text{ch}_2(T_X)=\text{ch}_2(V)-\left[C\right]
\label{anomaly1}
\end{equation}
Here, $\left[C\right]$ is the sum of the effective curve classes wrapped by the 5-branes in the compactification and $V$ is the $E_8 \times E_8$ gauge bundle.  As a conifold transition is traversed, $\text{ch}_2(T_X)$ changes. The second Chern character of the resolution geometry, $X_R$, is related to that of the deformation, $X_D$, as follows.
\begin{equation}
\text{ch}_2(T_{X_R})=\text{ch}_2(T_{X_D})+\left[\mathbb{P}^{1}s\right]
\label{ch2_change}
\end{equation}
Here the class $\left[\mathbb{P}^{1}s\right]$ is that of the exceptional locus of the transition. Given this change to the left hand side of (\ref{anomaly1}) across a conifold transition, we must include branes or bundles in the compactifications such that the right hand side of that equation changes in a consistent manner. The bridging curves, ${ C}_D \subset X_D$ on the deformation and ${ C}_R\subset X_R$ on the resolution, have classes that are related as follows.
\begin{eqnarray}
\left[ C_R \right] = \left[C_D\right] -\left[\mathbb{P}^{1}s\right]
\end{eqnarray}
As such, if these are the only parts of the gauge bundle and 5-brane sector that do not `spectate' through the geometric transition, then they exactly balance the change (\ref{ch2_change}) in the second Chern classes of the tangent bundle on the left hand side of (\ref{anomaly1}) ensuring that this anomaly cancelation condition is satisfied on both sides of the transition.

It was shown in \cite{Anderson:2022bpo} that, if one considers a heterotic background consisting only of bridging NS5-branes and 5-branes that spectate across the transition, then the total degrees of freedom of the theory (including the vector and chiral multiplets) are preserved across the conifold transition. In particular the massless singlets of the 5-brane-only theory, as counted by
\begin{equation}
h^{1,1}(X_D)+h^{2,1}(X_D)+h^0(C_D,{\cal N}_{C_D}) = h^{1,1}(X_R)+h^{2,1}(X_R)+h^0(C_R,{{\cal N}}_{C_R})\;,
\end{equation}
agree perfectly despite the fact that the Hodge numbers of the Calabi-Yau 3-folds are changing. In the above formula ${\cal N}_{C_D}$ and ${{\cal N}}_{C_R}$ are the normal bundles to $C_D$ and $C_R$ respectively.

The heterotic bridging brane theories described above readily extend to linked perturbative solutions of the theory by small instanton transitions \cite{Witten:1995gx,Ganor:1996mu,Ovrut:2000qi,Buchbinder:2002ji} which transform the NS5-branes described above back into slope-stable holomorphic vector bundles (via a so-called Hecke Transform \cite{Ovrut:2000qi}). This allows for an identification of pairs of Calabi-Yau 3-folds and vector bundles $(X_D, V_D)$ and $(X_R, {V_R})$ linked by conifold transitions of the Calabi-Yau base manifolds. In this context, we observed that, once again, the net numbers of massless singlets of the theories agree exactly.
\begin{equation}
h^{1,1}(X_D)+h^{2,1}(X_D)+h^1(X,\text{End}_0(V_D))=h^{1,1}({X_R})+h^{2,1}({X_R})+h^1({X_R},\text{End}_0({V_R})) 
\label{singlet_match1}
\end{equation}
In addition, the multiplicity of charged matter is also identical. As a special case of these examples, our results for smooth bundles reproduce known examples of pairs of heterotic geometries that arise from $(0,2)$ target space duality \cite{Distler:1995bc,Blumenhagen:1997vt,Blumenhagen:1997cn,Blumenhagen:2011sq,Rahn:2011jw,Anderson:2016byt} in which two distinct manifold bundle pairs share a non-geometric phase of a GLSM.

\vspace{0.2cm}

In this work, our goal is to move beyond the matching of the massless spectrum described above and probe the structure of the pairs of effective theories. The fact that the background heterotic geometries in \cite{Anderson:2022bpo} are linked by geometric transitions tells us that at minimum the four dimensional theories should be connected at some points in field space. As a result, one way to phrase the central question of this work is to ask whether or not the correspondence we find between complete manifold and brane/bundle geometries (and associated 4-dimensional physics) corresponds to a branch change in the vacuum space of the theory or to a true duality? The agreement in dimension of the moduli space suggests the latter, but this must be checked.

\begin{figure}[!t]\centering
\includegraphics[width=1\textwidth]{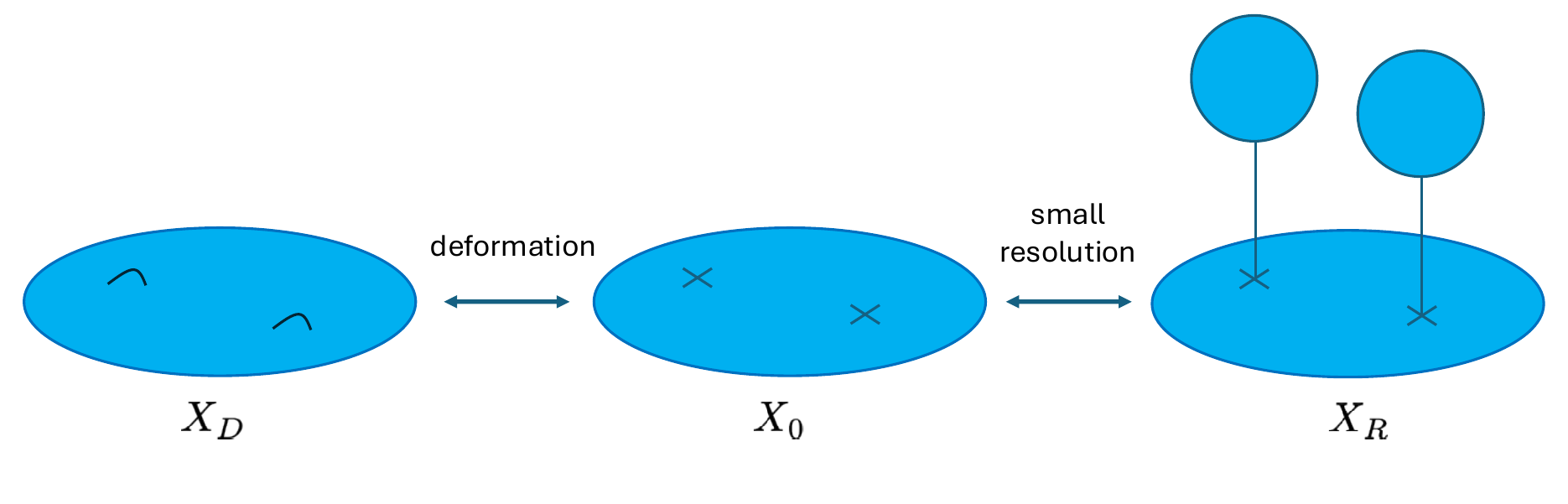}
\caption{{\it In a Calabi-Yau conifold transition, two smooth Calabi-Yau manifolds (denoted $X_D$ for deformation and $X_R$ for resolution) are connected by a shared singular variety, $X_0$, with nodal singularities. From $X_0$ the singularities can be smoothed by deformation of complex structure (corresponding to increasing non-trivial $S^3$ cycles from zero size) to obtain $X_D$.  A small resolution of the singular geometry (by replacing the singular points with curves that are topologically $\mathbb{P}^1$s) results in $X_R$.}}
\label{conifold_phil}
\end{figure}

To put these distinct possibilities in a familiar context, there are several examples of geometric correspondences of both types mentioned above in $4$-dimensional, ${\cal N}=2$ theories. A conifold transition between two Calabi-Yau 3-folds, as shown in Figure \ref{conifold_phil} is a clear example of vacuum branch change in the ${\cal N}=2$ context. Traversing the transition corresponds to moving between the Higgs and Coulomb branches of the associated 4-dimensional theory \cite{Strominger:1995cz,Greene:1995hu}. In this case the vacuum spaces are connected, corresponding to the conifold limit itself in the 3-fold geometry, but the effective physics is different across the resolution and deformation sides of the conifold. This follows from the physical interpretation of the distinct topological data (Hodge numbers, etc), in the two Calabi-Yau 3-folds. This branch change arises because of a true singularity in the apparent field space of the 4-dimensional effective theory which is repaired by the addition of extra light states at the conifold limit \cite{Strominger:1995cz,Greene:1995hu}.

In contrast to this, other geometric correspondences do lead to true dualities of ${\cal N}=2$ theories. Examples of these dualities include mirror symmetry of Calabi-Yau 3-folds \cite{Candelas:1990rm} and flop transitions in Type IIB theories \cite{Aspinwall:1993nu}. In these cases, the full moduli space of the 4-dimensional, ${\cal N}=2$ theory is visible from compactification on each of the underlying geometries: i.e. the Calabi-Yau 3-fold and its mirror or two different K\"ahler small resolutions of the same singular variety, whose field theories can be related by an analytic continuation of a K\"ahler modulus $t \rightarrow -t$. In the case of flop transitions, although the Calabi-Yau 3-fold geometry becomes singular, the field space singularity (and extra light states) associated to the conifold limit can be avoided by the choice of a non-vanishing vev for a K\"ahler axion, leading to a smooth field space connection between the two flop ``phases" \cite{Aspinwall:1993nu}.

We will test whether the heterotic theories associated to conifold transitions with bridging branes are in fact dual by providing a (partial) mapping of 4-dimensional fields. We will then compare the form of the effective field theories of the two smooth heterotic compactifications. With this proposed mapping we find that the gauge kinetic functions and Yukawa couplings in the holomorphic superpotentials of the two theories agree exactly. This latter result is strong evidence that the correspondence we're studying here is indeed a new, truly ${\cal N}=1$, duality and not merely a branch change in the effective theory. For example, in the case we present in detail, the comparison of the perturbative Yukawa couplings alone corresponds to the exact matching of $147,440$ distinct holomorphic functions of the 4-dimensional fields between the two theories.

The structure of the paper is as follows. In the rest of this section we review some of the results from \cite{Anderson:2022bpo} that will be required in our analysis. In Section \ref{mappymcmapface} we outline a proposal for the mapping of the four-dimensional moduli fields across the putative duality. In Section \ref{branemap} this proposal is laid out in detail and is shown to be compatible with the structure of the infinitesimal moduli deformations on either side of the transition for the case of a pair of heterotic NS5 brane systems. A similar analysis is performed for a pair of heterotic compactifications with non-trivial gauge bundles in Section \ref{mrbun}. In Section \ref{yukkysection} we provide the strongest evidence in the paper for the proposed ${\cal N}=1$ duality by showing that the moduli map put forward in previous sections leads to agreement of the perturbative superpotential Yukawa couplings as functions of the four dimensional fields. Finally, in Section \ref{concsec} we provide a summary of our results and describe some future directions of research.

\subsection{Calabi-Yau conifold transitions}\label{sec:conifolds}
We begin by reviewing a few basic facts about Calabi-Yau 3-fold conifold transitions. As described above, conifold transitions describe two smooth Calabi-Yau 3-fold varieties, connected at a singular shared locus in their moduli spaces. This singular locus is reached by a tuning of the complex structure of a smooth manifold $X_D$ on the so-called ``deformation" side of the conifold (a collapsing of 3 (real)-dimensional cycles). The same locus is reached from the moduli space of another smooth manifold, $X_R$, referred to as the ``resolution" geometry, by a tuning of a K\"ahler modulus to zero (i.e. a shrinking of complex curves). If $X_0$ is the singular (nodal) variety at which these two geometries meet, we say that $X_D$ is a smoothing, or deformation, of the singular locus, while $X_R$ is a small resolution. We’ll consider here the case where a single class of $\mathbb{P}^1$s, i.e. a single K\"ahler modulus, is shrinking to zero size on the resolution side.

It is well known that the Hodge and Euler numbers of the resolution are defined in terms of those of the deformation via \cite{Candelas:1987kf}
\begin{align}
&h^{1,1}(X_R)=h^{1,1}(X_D)+1 \;, \\
&\chi(X_R)=\chi(X_D)+2N \;.
\end{align}
Here $N$ is the number of nodal points in $X_0$. Moreover, in \cite{Anderson:2022bpo} we observed that 
\begin{align} \label{p1c2}
&\text{ch}_2(T_{X_R})=\text{ch}_2(T_{X_D})+\left[\mathbb{P}^{1}\right] \\
&0 \to f^*(\Omega_{X_0}) \to \Omega_{X_{R}} \to {\mathcal O}_{\mathbb{P}^1\mathrm{s}}(-2) \to 0  \label{hecke_tan}
\end{align}
where $\left[\mathbb{P}^1\right]$ denotes the curve class of the family of shrinking $\mathbb{P}^1$s, and $f\colon X_R \to X_0$ the small contraction map. Above, the $\Omega$'s denote the holomorphic cotangent sheaves, and $\mathcal{O}_{\mathbb{P}^1\mathrm{s}}$ is the skyscraper sheaf associated to the exceptional $\mathbb{P}^1s$. The geometric program of \cite{Anderson:2022bpo} utilized \eref{hecke_tan} to demonstrate that the change to the holomorphic cotangent bundle is similar in form to that of a so-called `Hecke Transform' \cite{Ovrut:2000qi} appearing in heterotic small instanton transitions \cite{Witten:1995gx,Ovrut:2000qi,Buchbinder:2002ji}.

One important note is that small resolution geometries always come in pairs due to the fact that an orientation must be chosen for the $\mathbb{P}^1s$. The two resolution geometries (i.e. $X_{R_1}$ and $X_{R_2}$) are related by a flop transition. To give a concrete illustration of such Calabi-Yau geometries, consider the well known quintic 3-fold in complex projective 4-space which we'll denote as $X_D=\mathbb{P}^4[5]$ (with Hodge numbers $h^{1,1}(X)=1,h^{2,1}(X)=101$). Tuning the complex structure of this manifold can take the defining quintic equation to the following nodal form.
\begin{equation}
l_1^{(1)}(y)q_4^{(2)}(y)-l_1^{(2)}(y)q_4^{(1)}(y)=0 \label{nodal1}
\end{equation}
where $l_1^{(i)}$ and $q_4^{(i)}$ are linear and quartic polynomials respectively in the homogeneous coordinates $y_i,,~i=0,\ldots 4$, of $\mathbb{P}^4$. The resulting Calabi-Yau  3-fold defined by \eref{nodal1} is singular at 16 points (given algebraically by $l_1^{(1)}=l_1^{(2)}=q_4^{(1)}=q_4^{(2)}=0$). The singular geometry of \eref{nodal1} can be repaired by a small resolution -- in which the 16 singular points are replaced\footnote{Note that for this nodal quintic and its small resolution, the 16 $\mathbb{P}^1s$ are all in the same class in the Mori cone of the resolution geometry.} by $\mathbb{P}^1s$. Such small resolutions come in pairs (associated to a choice of orientation in the $\mathbb{P}^1$s replacing the singular points). One such smooth manifold is given as a complete intersection in a product of projective spaces (CICY),
\begin{equation}\label{small_res1}
X_{R_1} =
\left[
\begin{array}{c | c c}
\mathbb{P}^1 & 1 & 1 \\
\mathbb{P}^4 & 1 & 4 
\end{array}
\right] \,,
\end{equation}
with Hodge numbers $h^{1,1}(X)=2$ and $h^{2,1}(X)=86$ and defining equations
\begin{align}
&x_0l_1^{(1)}(y)+x_1l_1^{(2)}(y)=0 \\
&x_0q_4^{(1)}(y)+x_1q_4^{(2)}(y)=0
\end{align}
where $x_0,x_1$ are the coordinates of the ambient $\mathbb{P}^1$.

 The second small resolution of the nodal quintic is related to $X_{R_1}$ by a flop transition and can be described as a co-dimension 2 complete intersection in a 5-dimensional toric ambient space. We will denote this manifold by its toric weight matrix (i.e. GLSM charge matrix) as follows.
\begin{eqnarray} \label{small_res2}
X_{R_2} &=& \left[\begin{array}{ccccccc|cc} y_0&y_1&y_2&y_3&y_4&x_0&x_1&P_{1,4}^1&P_{1,4}^2\\ 0&0&0&0&0&1&1&1&1\\ 1&1&1&1&1&3&0&4&4\end{array} \right] ~~.
\end{eqnarray}
The manifold $X_{R_2}$ is birational to $X_{R_1}$ above, with the same Hodge numbers and defining equations
\begin{align}
&x_0l_1^{(1)}(y)+x_1q_4^{(1)}(y)=0 \\
&x_0l_1^{(2)}(y)+x_1q_4^{(2)}(y)=0
\end{align}

\vspace{0.1cm}
With these preliminary results in hand, we turn now to a review of the underlying geometry of curves and branes described in \cite{Anderson:2022bpo} which plays a crucial role in conifold transitions in ${\cal N}=1$ theories.

\subsection{Bridging branes and Calabi-Yau geometric transitions}\label{sec:bridge_branes}

The correspondences between 4-dimensional theories observed in \cite{Anderson:2022bpo} are rooted in several observations about the geometry of Calabi-Yau conifold transitions. As noted in \cite{Anderson:2022bpo}, every Calabi-Yau conifold transition comes equipped with a set of curves/divisors that are intrinsic to the transition and appear as Weil, non-Cartier divisors in the singular limit. Due to the special role that these curves play in connecting two geometries leading to seemingly dual theories, we will refer to these curves as \emph{bridging curves} and the branes that wrap them as \emph{bridging branes}.

To define these, consider the following nodal Calabi-Yau manifold defined in a compact, complex ambient space ${\cal A}$ (with coordinates $y_i$).
\begin{equation}
p_{nodal}=f_1(y)g_2(y)-f_2(y)g_1(y)=0~\label{nodal}
\end{equation}
In such a nodal geometry, there are 4 special divisors defined by systems of equations of the form 
\begin{equation}\label{wnc}
f_1=f_2=0 \;\;,\;\; g_1=g_2=0 \;\;, \;\; f_1=g_1=0 \;\; \text{and} \;\; g_2=f_2=0\;.
\end{equation}
The solution to any set of equations given in (\ref{wnc}) automatically satisfies the nodal relation in \eref{nodal}. They thus combine with $p_{nodal}$ to form a non-complete intersection in ${\cal A}$ and hence define Weil, non-Cartier divisors in the nodal Calabi-Yau geometry. 

We can consider the fate of the subvarieties defined by \eref{wnc} upon desingularizing the nodal Calabi-Yau variety. We'll begin with a small deformation, that is, a change in complex structure. Upon deformation of the nodal Calabi-Yau variety to a smooth hypersurface in ${\cal A}$ we have the following.
\begin{equation}
p_{nodal}=0 \to p_{nodal}+\delta p=0
\end{equation}
The system of equations given in \eref{wnc} now becomes a \emph{complete intersection} in the Calabi-Yau and the dimension of the solution set changes. Thus, a divisor of the form above defined in the nodal limit becomes a smooth curve in the deformation manifold, $X_D$. 

We will illustrate bridging branes in the resolution geometry in the example of a case which can be described algebraically as a $\mathbb{P}^1$ split \cite{Candelas:1987kf}. In such a case, $X_R$ can be written schematically as,
\begin{align} \label{mrres}
&x_0f_1(y) + x_1 f_2(y)=0 \\ \nonumber
&x_0 g_1(y)+x_1 g_2(y)=0 
\end{align}
in ${\mathbb P}^1 \times {\cal A}$ where $\{x_0,x_1\}$ are the coordinates on the $\mathbb{P}^1$. Note that writing this system of equations in matrix form yields
\begin{equation}
{\cal M} \,\vec{X}=\left( \begin{array}{cc} f_1 & f_2\\
g_1 & g_2 \end{array} \right)\cdot \left( \begin{array}{c} x_0 \\ x_1 \end{array} \right) =0
\end{equation}
which has a solution if and only if $det({\cal M})=p_{nodal}=0$. 
In such a resolution geometry, the bridging curve is defined by a system of equations such as 
\begin{align} \label{r1}
&x_0=0 \;,\\  \label{r2}
&p=p_{nodal}+\delta p=0\;.
\end{align}
This system of equations defines a smooth curve in $X_R$ for generic defining equations $p=0$. However, in the limit that $\delta p \to 0$, (\ref{r1}) and (\ref{r2}) define a divisor in $X_R$ which approaches one of the Weil, non-Cartier divisor given in \eref{wnc} in the limit that the resolutions $\mathbb{P}^1$s shrink to zero volume.

As a concrete example, the bridging branes connecting the quintic, $\mathbb{P}^4[5]$, with the small resolution given in \eref{small_res1} are as follows.
\begin{center}
 \begin{tabular}{ |c|c|c|}
\hline
&Deformation &Resolution\\
\hline
& & \\
Bridging Curves & $l_1^{(1)}=0,~q_4^{(1)}=0$ & $x_0=0,~f_{0,5}=0$\\
& & \\
\hline
\end{tabular}~~
\end{center}
In this example, the curve classes are $[C_D]=4D^2$ and $[C_R]=5D_1D_2$ respectively, where $D$ denotes the single divisor class of the quintic and $D_1,D_2$ denote divisors formed by the restriction of the ambient hyperplane classes in \eref{small_res1}.

It should be noted that the nodal locus in \eref{nodal} is shared between the small resolution manifold given in (\ref{mrres}) and its geometric flop. In many cases, this geometric flop can be constructed as toric complete intersection in an ambient space which is a $\mathbb{P}^1$ bundle over ${\cal A}$. The analysis of bridging branes and their divisor limits in this case is very closely analogous to that presented above.

\vspace{0.1cm}

As was described around (\ref{p1c2}), if the correct bridging curves are chosen to link a deformation and resolution geometry, their class changes across the conifold transition in exactly the same fashion as the second Chern class of the manifold. Thus their presence can ensure that anomaly cancelation is preserved in heterotic string compactifications traversing the transition. For example, for the conifold pair given by the quintic threefold and the resolution in \eref{small_res1}
\begin{equation}
\text{ch}_2(T_{X_R})=-5D_1D_2-6 D_2^2=\text{ch}_2(T_{X_D})+[\mathbb{P}^1s]=-10D_2^2+ (-5D_1D_2+4D_2^2)
\end{equation}
with $D_1,D_2$ the restrictions of the two ambient hyperplane divisors. As expected, the class of the exceptional locus, $[\mathbb{P}^1s]=-5D_1D_2+4D_2^2$, is exactly the difference in the classes of the bridging curves given in the table above.

\subsection{Bridging curves and heterotic NS$5$-brane theories}\label{1st_brane_eg}
As described above, in the case of heterotic theories, one could consider backgrounds of the theory comprised of Calabi-Yau 3-folds (linked by a conifold transition) and NS5-branes wrapping the bridging curves. As an explicit example, let us consider the case of the $\mathbb{P}^1$-split of the quintic described earlier.
\begin{eqnarray} \label{spliteg1}
X_D = \left[\begin{array}{c|c} \mathbb{P}^4 & 5 \end{array}\right] \longleftrightarrow \left[\begin{array}{c|cc} \mathbb{P}^1&1&1 \\ \mathbb{P}^4 &1&4  \end{array} \right] =X_R
\end{eqnarray}
In this case the bridging curves are defined by the following normal bundles.
\begin{eqnarray} \label{bridgeceg}
{\cal N}_{C_D} = {\cal O}(1) \oplus {\cal O}(4) \;\;\;,\;\;\; {\cal N}_{C_R} = {\cal O}(1,0) \oplus {\cal O}(0,5)
\end{eqnarray}
To fully satisy the anomaly cancellation condition of \eref{anomaly1} it is necessary to add either a spectator bundle or a spectator 5-brane to these backgrounds which will saturate the anomaly cancellation condition. If we opt for the latter possibility one can choose to include such objects on curves with the following normal bundles on the two sides.
\begin{eqnarray}
{\cal N}_{C_{Ds}} = {\cal O}(2) \oplus {\cal O}(3) \;\;\;,\;\;\; {\cal N}_{{C}_{Rs}} = {\cal O}(0,2) \oplus {\cal O}(0,3)
\end{eqnarray}
These ``spectate" through the conifold transition in that they do not intersect the singular points and their curve class carries through unchanged. 

\begin{figure}[!t]\centering
\includegraphics[width=0.5\textwidth]{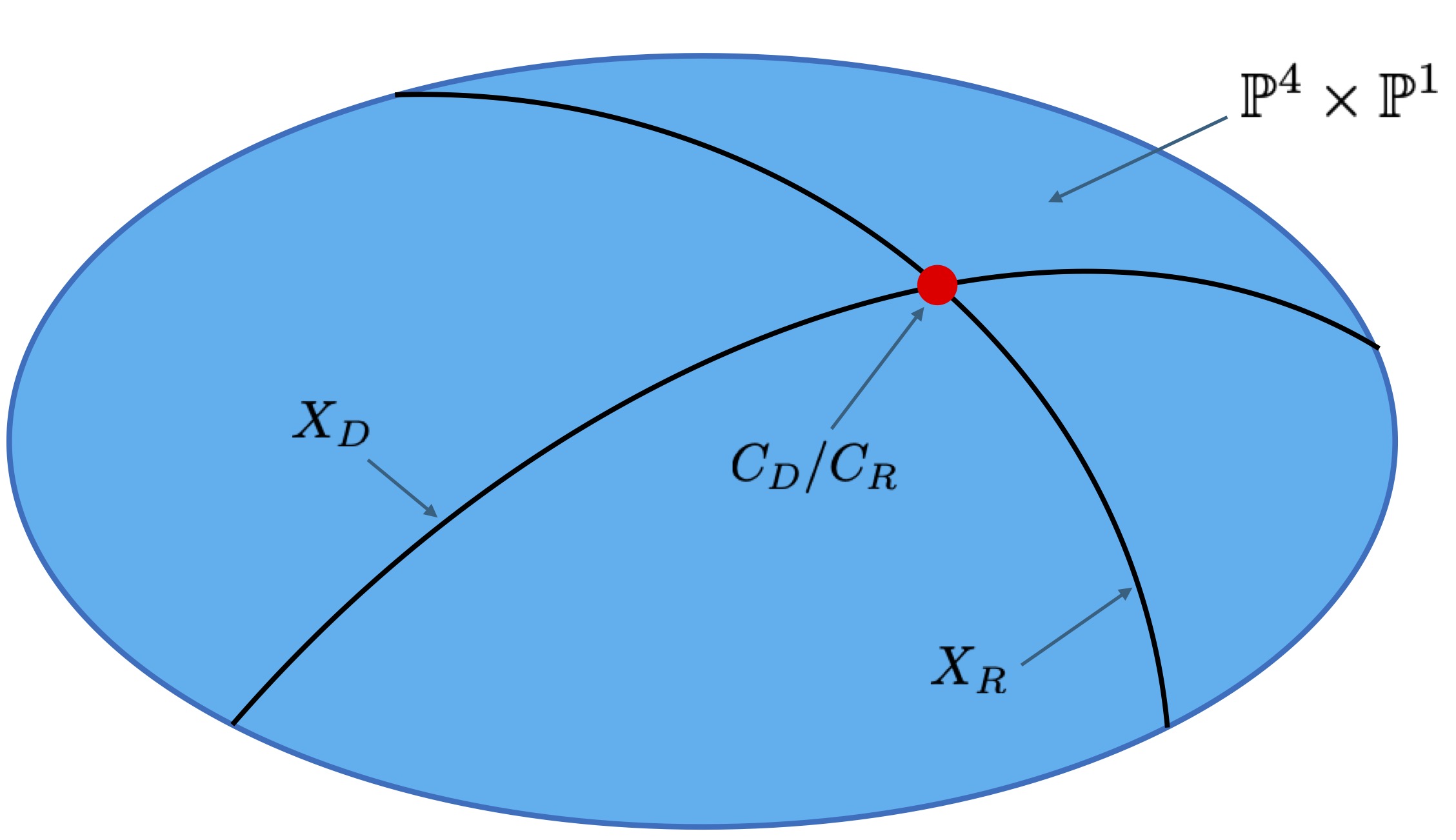}
\caption{{\it A schematic of the intersection of the deformation and resolution manifolds in a Calabi-Yau conifold transition, described in a shared ambient space.}}
\label{conifoldtrans}
\end{figure}

Counting the chiral multiplet moduli on both sides of this transition, we arrive at the following.
\begin{eqnarray} \label{matchingeg}
\begin{array}{|c|c|c|c|c|c|c|c|} \hline &&&&& \\[-1em] & h^0({\cal N}_{{C_D}/{C_R}})& h^0({\cal N}_{C_{Ds}/{C}_{Rs}}) & h^{1,1}(X_D/{X_R}) & h^{2,1}(X_D/{X_R})  & \text{Total} \\ \hline \text{Deformation} & 38&30&1&101&170 \\ \text{Resolution} & 52&30&2&86& 170\\\hline \end{array}
\end{eqnarray}
Thus we see that these moduli match as claimed. In addition, as shown in \cite{Anderson:2022bpo}, it is easy to see that the genus of the curves agree and hence, the vector multiplet spectra also match. To show this, consider rewriting the deformation side geometry as,
\begin{eqnarray}
X_D = \left[ \begin{array}{c|cc} \mathbb{P}^1 &1&0 \\\mathbb{P}^4 &0&5 \end{array} \right] \;\;\; \text{and} \;\;\; {\cal N}_{C_D} = {\cal O}(1,1) \oplus {\cal O}(1,4) \;.
\end{eqnarray}
The geometrical setup intrinsic to $X_D$ here is identical to the one above, it is simply embedded in a different ambient space. In this description it can be easily seen that $C_D$ and $C_R$ form the same curve in $\mathbb{P}^1 \times \mathbb{P}^4$ - which is the locus where the two Calabi-Yau varieties intersect in the common ambient space (see Figure \ref{conifoldtrans}). It is therefore manifest that any intrinsic geometric property of the two curves, such as genus, matches across the duality.

\subsection{Perturbative heterotic pairs from bridging branes and small instanton transitions}
Instead of utilizing the spectator branes described in the previous subsection, we could also complete the anomaly cancellation condition (\ref{anomaly1}) by choosing a ``spectator" gauge bundle. For the 5-brane theory above this bundle must satisfy $c_2(V_s)=6D^2$ on $\mathbb{P}^4[5]$. For example, one such choice is the bundle defined via the following short exact sequence.
\begin{equation}
0 \to {\cal O}(-4) \to {\cal O}(-1)^{\oplus 4} \to V_{Ds} \to 0
\end{equation}
This bundle can be merged with the 5-brane wrapping the bridging curve $C_D$, whose normal bundle is given in \eref{bridgeceg}, via a small instanton transition. More precisely, the 5-brane/small instanton can be described via an ideal sheaf as
\begin{equation}\label{ideal_sheaf_c}
0 \to {\cal O}(-5) \to {\cal O}(-1)\oplus {\cal O}(-4) \to {\cal I}_{C_D} \to 0 \;.
\end{equation}
Then the singular sheaf $\hat{V}=V_{Ds} \oplus {\cal I}_{C_D}$ arises naturally in a so-called Hecke transform describing the small instanton transition \cite{Ovrut:2000qi}.
\begin{equation}
0 \to \hat{V}_D \to V_{Ds}\oplus {\cal O}_X \to {\cal O}_{C_D} \to 0
\end{equation}
The singular sheaf $\hat{V}_D$ can be deformed into a smooth vector bundle defined by the following short exact sequence.
\begin{equation}
0 \to 0 \to {\cal O}(-5)\oplus{\cal O}(-4) \to {\cal O}(-1)^{\oplus 5}\oplus {\cal O}(-4) \to V_D \to 0~.
\end{equation}
This bundle is equivalent to one for which the factor of ${\cal O}(-4)$ is omitted in the first and second entries in the short exact sequence above. In this latter form the bundle is a familiar one, a deformation of the holomorphic cotangent bundle.
\begin{equation} \label{buneg1}
0 \to 0 \to {\cal O}(-5) \to {\cal O}(-1)^{\oplus 5}\to V_D \to 0~
\end{equation}

Turning now to the the resolution geometry given in \eref{spliteg1}, we can analogously extend the 5-brane solution to that of a smooth bundle. Here we define the spectator bundle as
\begin{equation}
0 \to {\cal O}(0,-4) \to {\cal O}(0,-1)^{\oplus 4} \to {V}_{Rs} \to 0~.
\end{equation}
The ideal sheaf associated to the bridging brane that matches that with the ideal sheaf given in \eref{ideal_sheaf_c} in the shared nodal limit is given by
\begin{equation}
0 \to {\cal O}(-1,-5) \to {\cal O}(-1,0)\oplus {\cal O}(0,-5) \to {\cal I}_{C_R} \to 0 \;.
\end{equation}
Performing a small instanton transition to absorb the 5-brane wrapping this bridging curve into ${V}_{Rs}$ leads to the following smooth bundle.
\begin{equation} \label{buneg2}
0 \to {\cal O}(-1,5)\oplus {\cal O}(0,4) \to {\cal O}(0,-5)\oplus {\cal O}(-1,0)\oplus{\cal O}(0,-1)^{\oplus 4} \to V_R \to 0
\end{equation}

As discussed in detail in \cite{Anderson:2022bpo} the massless spectra of such heterotic theories will agree. In this instance the massless singlet spectrum in the two linked theories is as follows.
\begin{eqnarray}\label{bundle_match}
\begin{array}{|c|c|c|c|c|c|c|} \hline &&&& \\[-1em] & h^0(\text{End}_{0}(V_D)/\text{End}_{0}(V_R)) & h^{1,1}(X_D/X_R) & h^{2,1}(X_D/X_R)  & \text{Total} \\ \hline \text{Deformation} & 325&1&101&427 \\ \text{Resolution} & 339&2&86& 427\\\hline \end{array}
\end{eqnarray}

These examples exhibit the intriguing structure of these potentially dual heterotic theories. We now turn to studying whether such seemingly different geometric backgrounds really do lead to dual 4-dimensional theories.

\section{Towards a moduli map between dual theories} \label{mappymcmapface}
The correspondence of spectra in \eref{matchingeg} and \eref{bundle_match} hints that, if the corresponding 4-dimensional theories are indeed dual, then the duality must have a number of novel features. The 4-dimensional fields corresponding to geometric K\"ahler, complex structure, and bundle moduli, appear in the heterotic effective theory in distinct ways. For example, the holomorphic superpotential is known to be independent of K\"ahler moduli in perturbative Calabi-Yau compactifications \cite{Green:1987mn}. Moreover, the non-perturbative contributions to the superpotential are known to have a specific functional form in terms of these variables \cite{Dine:1986zy,Dine:1987bq}. Equally, the gauge kinetic function is known in perturbative vacua to be a function of the K\"ahler moduli and dilaton only \cite{Lukas:1997fg}. How then can two theories be dual when the duality would seemingly map these degrees of freedom into one another?

To directly answer this question, one needs a putative mapping of fields that could be explicitly tested. We are limited by the fact that many aspects of the 4-dimensional EFT remain unknown for ${\cal N}=1$ theories (for example the form of the matter field K\"ahler metric). In addition, neither the underlying stringy correspondence or a field-theoretic description of the common point in moduli space are yet understood. However, as we will demonstrate in the following sections, there is information that can be extracted from the potentially ``dual" manifold, brane and bundle geometries themselves. The infinitesimal geometric moduli of the full heterotic background are defined by highly restrictive equivalence classes which contain information that constrains any purported matching of 4-dimensional fields. We will use this structure to form a (partial) conjectural mapping of these degrees of freedom that we can then test using known aspects of the EFT, including the holomorphic superpotential.

In more detail, we will base the correspondence on the following two requirements:
\begin{itemize}
\item Physical fluctuations of the background geometries (i.e. the manifolds, branes and bundles) should map to physical fluctuations.
\item The two geometries, and their associated theories, should ``meet in the middle" at the conifold locus.
\end{itemize}
As we will observe below, the two conditions above impose strong constraints on the form of any purported duality between theories. However, these conditions are not enough to uniquely determine the moduli map (especially in the K\"ahler sector of the theory) and we will carefully phrase the remaining ambiguity in the mapping below. We begin by considering the most direct observations we can make regarding the structure of the two 4-dimensional EFTs that we hope to compare.

\subsection{Gauge kinetic functions}

Some of the most accessible parts of the four dimensional effective theory of a heterotic compactification are the gauge kinetic functions. These quantities are holomorphic, one loop exact and entirely determined by topological properties of the manifold and bundle. The functional form of the gauge kinetic functions is so simple that they allow us to identify part of the field space map across our putative duality immediately. We therefore examine this information here first, before proceeding to consider the field space map in more generality in the next subsection.

The gauge kinetic functions of the two gauge groups in a heterotic theory are given, in the absence of M5 branes, by
\begin{eqnarray} \label{simpcase}
f^{\pm} = S \pm \beta_i T^i \;,
\end{eqnarray}
where $S$ is the dilaton and the $T^i$ are the K\"ahler moduli. Here we have
\begin{eqnarray} \label{beta1}
\beta_k = \frac{\epsilon_S}{v^{\frac{1}{3}}} \int \omega_k \wedge \frac{1}{16 \pi^2} \left( \tr F \wedge F - \frac{1}{2} \tr R \wedge R\right) \;,
\end{eqnarray}
where $v$ is the coordinate volume of the Calabi-Yau 3-fold, $\epsilon_S$ is the usual coupling parameter of the strongly coupled description \cite{Lukas:1998hk}, $F$ is the field strength of one of the $E_8$ bundles, and the $\omega_k$ are a basis of (1,1)-forms. 

Since we know that the $E_8$ sectors with non-trivial bundles have matching spectrum for our putative dual pairs, if the phenomenon is to be a true duality then their gauge kinetic functions must be equal at correctly related points in field space. Therefore matching these gauge kinetic functions, for example in the combinations $f^+ \pm f^-$,  gives us a matching of two moduli across the duality, namely the dilaton and one linear combination of the K\"ahler moduli.
\begin{equation} \label{part_mod_match}
\begin{array}{rcl}
S_D  &\leftrightarrow& S_R \\
\text{ch}_{2,i}(TX_D)\,T_D^i &\leftrightarrow& \text{ch}_{2,i}(TX_R)\,T_R^i
\end{array}
\end{equation}
Here we have written $\text{ch}_{2,i}(TX)$ as shorthand for $\int \omega_i \wedge \text{ch}_2(TX)$.

As an example of this relationship between fields consider the case given in (\ref{spliteg1}), (\ref{buneg1}) and (\ref{buneg2}). In this example, (\ref{part_mod_match}) leads to the following linear relationship amongst K\"ahler moduli across the duality, in addition to the equivalence of the four dimensional dilatons.
\begin{eqnarray}
50 T_D = 24 T^0_R + 50 T^1_R
\end{eqnarray}
Here we have not put a superscript on the deformation side K\"ahler modulus because there is only one such field. In addition, we have chosen to label the `extra' K\"ahler modulus on the resolution side, which is set to zero in the nodal limit, as $T^0$.

\vspace{0.2cm}

More generally, we can consider the case where we do have M5 branes present in vacuum. The explicit formulae for the gauge kinetic function in such situations is as follows \cite{Brandle:2003uya}.
\begin{eqnarray} \label{fullcase}
    f^{1} &=& S + \beta^{(1)}_iT^i \\ \nonumber
    f^{2} &=& S + \beta^{(2)}_iT^i-2 \sum_a Z^a
\end{eqnarray}
Here the $\beta^{(i)}$ are defined as in (\ref{beta1}) with the relevant field strength being used for each $E_8$ factor. The sum in $f^2$ runs over all M5 branes in the vacuum. The precise relation of the fields $Z^a$ to quantities in the dimensional reduction is complicated, but at essence they are degrees of freedom associated with M5 brane position moduli. In this more general setting we see that one does not obtain equal dilatons across the dual pair. Instead we are simply given two linear relations between superfields which involve more degrees of freedom than in the previous case. Note that, due to the integrability condition on the heterotic Bianchi Identity $\beta^{(1)} = -\beta^{(2)}$ in the case where the number of five branes is zero, and thus we recover (\ref{simpcase}) from (\ref{fullcase}) correctly.

\subsection{Mapping the 4-dimensional moduli fields}

In this section we will provide an intuitive, but schematic, overview of the field mapping we propose in the example of the type of manifold and brane system discussed in Section \ref{1st_brane_eg}. This proposal will then be made more explicit in the more detailed sections which follow.

Intuitively, the field mapping that we propose is based on the structure of bridging branes seen in Section \ref{sec:bridge_branes} above. In the notation of Section \ref{sec:conifolds} for conifolds described as $\mathbb{P}^1$-splits in CICYs, the conifold transition connects two manifolds
 \begin{eqnarray}
X_D = \left[\begin{array}{c|cc} \mathbb{P}^1&1&0 \\ {\cal A} &0& f  \end{array} \right]\longleftrightarrow \left[\begin{array}{c|cc} \mathbb{P}^1&1&1 \\ {\cal A} &p& q  \end{array} \right] =X_R \\
 \end{eqnarray}
 where $deg(f)=deg(p)+deg(q)$. Here, we have restricted our attention to a hypersurface on the deformation geometry for notational simplicity. In this setting, the polynomial defining relations of the manifolds and bridging branes of Section \ref{sec:bridge_branes} take the following form.
 
\begin{center}
 \begin{tabular}{ |c|c|c|} 
\hline
&Deformation &Resolution\\
\hline
& & \\
Manifold & $P_{1,0}^{(D)}, P_{0,f}^{(D)}$ & $P_{1,p}^{(R)},P_{1,q}^{(R)}$\\
& & \\
\hline
& & \\
Brane & $P_{1,p}^{(D)},P_{1,q}^{(D)}$ & $P_{1,0}^{(R)}, P_{0,f}^{(R)}$\\
& & \\
\hline
\end{tabular}
\end{center}
In this table, the subscripts denote the multidegrees of the polynomials $P$. From this table, it is clear that by choosing correlated polynomials as defining data for the manifold/brane, the bridging branes could sweep out literally the same locus within the ambient space (See Figure \ref{conifoldtrans}). This  hints towards a broad expectation that the role of the underlying polynomials should be simply switched in the two theories. That is, the defining equations that swept out a Calabi-Yau complete intersection manifold on one side of the conifold are the same algebraic expressions as those used to define a bridging curve on the other (and vice versa). It is worth noting that this is exactly the correspondence that arises in those dual theories which can be realized within the context of target space duality of $(0,2)$ GLSMs\footnote{In the GLSM context the polynomial equations that are being interchanged are those defining monad bundles and manifolds, respectively. This interchange can be related to the one alluded to above by small instanton transitions \cite{Anderson:2022bpo}}. However, as we will see below this naive ``polynomial interchange" is not quite correct. Instead, we will postulate a more refined correlation of the geometric degrees of freedom. In particular, a naive reading of the table above might lead one to think that the field mapping simply exchanged complex structure moduli with brane moduli (and vice versa) across the geometric transition. As we will sketch below, the actual field mapping is more intricate and depends crucially on the K\"ahler moduli of the theory, which are not included in the table above.

In order to get an intuitive feel for the proposed field mapping laid out in detail in the following sections, it is useful to group the 4D fields arising on the deformation side of the conifold in the following way.

\begin{eqnarray}
\begin{array}{|c|} \hline
 \text{Deformation~Fields}  \\ \hline \hline
 \text{``Spectator"~Complex~Structure~Moduli}  \\ \hline
 \text{``Active"~Complex~Structure~Moduli}\\ \hline
 \text{Brane~Moduli} \\ \hline
 \text{K\"ahler~Moduli} \\ 
 \hline
 \end{array}
 \end{eqnarray}
Here we refer to the complex structure related to collapsing $S^3$-cycles in the deformation geometry as ``active" and those degrees of freedom which survive to become moduli of the the nodal variety as ``spectator" complex structure. In analogous notation, on the resolution geometry, we will sub-divide the K\"ahler moduli space, splitting off the ``new" K\"ahler class (here we consider the case where $h^{1,1}$ changes by exactly one in the conifold transition).
\begin{eqnarray}
\begin{array}{|c|} \hline
 \text{Resolution~Fields}  \\ \hline \hline
 \text{``Spectator"~K\"ahler~moduli}  \\ \hline
 \text{``Active"~K\"ahler~Modulus}~({T_R}^0)\\ \hline
 \text{Brane~Moduli} \\ \hline
 \text{Complex~Structure~Moduli} \\ 
 \hline
 \end{array}
 \end{eqnarray}
Similarly to the complex structure moduli of the deformation side, the `spectator' K\"ahler moduli are simply those that survive into the nodal limit. In both cases, this is not meant to imply that these fields do not undergo a non-trivial mapping, but instead that the requirement that two geometries ``meet in the middle" imposes restrictions on the field mapping. In the case of the resolution geometry, in the limit that the new K\"ahler modulus vanishes, i.e. ${T_R}^0 \to 0$, the remaining K\"ahler moduli of the resolution must agree with those of the deformation in the nodal limit. That is
\begin{equation} \label{boatymcequationface}
T_D^a = T_R^a+f^a({T_R}^0)
\end{equation}
for some functions $f^a$ of ${T_R}^0$ (and possibly of other fields as well) such that $f^a(0)=0$.

In mirror symmetry for ${\cal N}=2$ theories, the complex structure degrees of freedom are mapped into K\"ahler moduli across a dual pair. In our current 4D ${\cal N}=1$ context, the complex structure, K\"ahler and brane moduli {\it all} experience an interchange. Importantly, the intrinsically ${\cal N}=1$ moduli (i.e. those of the branes in the heterotic context) are playing a crucial role in the geometric transition and in balancing the of degrees of freedom across the duality. The schematic mapping of fields can be described by the table below.
\begin{eqnarray}
\begin{array}{|c|c|} \hline
 \text{Deformation} &   \text{Resolution} \\ \hline
\text{Active~Complex~Structure~Moduli}&  \text{Brane~ Moduli},~\text{Active~K\"ahler~Modulus}\;\;{T_R}^{0} \\
\text{Spectator~Complex~Structure~Moduli} &  \text{Complex~Structure~Moduli}\\
\text{Brane~Moduli} &  \text{Complex~Structure~Moduli} \\
\text{K\"ahler~Moduli }\;\;T_R^a &   \text{Spectator~K\"ahler~Moduli}  \;\;{T_R}^a (a \neq 0) \\
 \hline
 \end{array} \label{moduli_match_table}
 \end{eqnarray}

The outline of the field mapping in \eref{moduli_match_table} is meant only to provide a coarse intuitive overview. To define the true (partial) mapping of fields, and justify the claimed interchange of degrees of freedom, we turn now to a detailed analysis of the manifold, brane, and in some cases bundle, geometries in consideration.

\section{The moduli map in a brane example} \label{branemap}

Let us illustrate the discussion in the case of a simple example. On the deformation side of the transition we shall consider the quintic Calabi-Yau manifold, with an NS5 brane wrapped on a specific curve.
\begin{eqnarray} \label{one}
X_D &=& \left[ \begin{array}{ccccccc|cc}y_0&y_1&y_2&y_3&y_4&x_0&x_1&P_{1,3}&P_{0,5}\\ 0&0&0&0&0&1&1&1&0 \\ 1&1&1&1&1&3&0&3&5 \end{array}\right] \\ \nonumber
{\cal N}_{{\cal C}_D} &=& {\cal O}_{X_D}(1,4)^{\oplus 2}
\end{eqnarray}
Here, $X_D$ is a GLSM charge matrix providing a description of the quintic and ${\cal N}_{{\cal C}_D}$ is the normal bundle to the five-brane stack. This model must be completed by a spectator brane or bundle to satisfy anomaly cancellation, but since this spectator structure is essentially completely unchanged by the transition we will not discuss it further here.

On the resolution side of the transition we have the following.
\begin{eqnarray} \label{two}
X_R &=& \left[\begin{array}{ccccccc|cc} y_0&y_1&y_2&y_3&y_4&x_0&x_1&P_{1,4}^1&P_{1,4}^2\\ 0&0&0&0&0&1&1&1&1\\ 1&1&1&1&1&3&0&4&4\end{array} \right]    \\ \nonumber
{\cal N}_{{\cal C}_R} &=& {\cal O}_{X_R}(1,3) \oplus {\cal O}_{X_R}(0,5)
\end{eqnarray}

How does the duality map between these two compactifications manifest itself at the level of the data required to define the vacua? First note that both configurations are defined by a set of polynomials of degrees $(1,3)$, $(0,5)$, $(1,4)$ and $(1,4)$ in terms of the same ambient space variables. It is tempting, therefore, to guess that the duality just maps configurations on the deformation with a given set of explicit choices of those polynomials to configurations on the resolution with exactly the same defining data.

This is almost, but not quite correct. The issue with such a straightforward matching of degrees of freedom can be seen in considering perturbations to the defining data which is physical on one side of the duality but not on the other. For example, let us for simplicity, by use of a coordinate transformation, set the defining relation of degree $(1,3)$ to be simply $P_{1,3}=x_0$. This would be a defining equation of the manifold on the deformation and a defining equation of the bundle on the resolution if the simple proposal of the proceeding paragraph were to be correct. Now consider making a change to the degree $(1,4)$ polynomials of the following form.
\begin{eqnarray}
\delta P^{\alpha}_{1,4} \propto P_{1,3} \,l^{\alpha}_{0,1}
\end{eqnarray}
Here $\alpha=1,2$ and the $l^{\alpha}_{0,1}$ are arbitrary functions of degree $(0,1)$. Such a change in the $(1,4)$ polynomials constitutes a physical change in the vacuum on the resolution, but not on the deformation, because the defining relations of $X_D$ mean that this change in the defining equations of the brane goes to zero on the Calabi-Yau. Similarly, changing the degree $(0,5)$ polynomial on the resolution by a term proportional to the nodal quintic constructed from the two $(1,4)$ polynomials does not result in a physical change in the vacuum, whereas changing the quintic polynomial in such a fashion on the deformation most certainly does.

This issue can be fixed by correlating the structure of some of the defining polynomial data involved in constructing these compactifications. In particular, let us without loss of generality write the degree $(1,4)$ polynomials as follows.
\begin{eqnarray} \label{14s}
P_{1,4}^{\alpha} = l^{\alpha} x_0 + q^{\alpha} x_1
\end{eqnarray}
We then demand that the polynomial long division of the quintic polynomial by $q^{\beta}$ is precisely $\epsilon_{ \alpha \beta} l^{\alpha}$.
\begin{eqnarray} \label{p5split}
P_{0,5} = P'_{0,5} + \epsilon_{\alpha \beta}  l^ {\alpha} q^{\beta}
\end{eqnarray}
Here $P_{0,5}'$ is the remainder of the long division by $q^{\beta}$. As we will see, correlating the defining data of the compactifications in this way prevents physical fluctuations in one theory being associated to irrelevant changes in defining data in the other. Before getting into these details however, we should make a few points about the above correlation of data. Firstly, although we have specialized the defining data above so that the degree $(1,4)$ and degree $(0,5)$ polynomials are correlated, we can still describe a completely general vacuum in this way on both the deformation and resolution sides of the duality. On the deformation side, if $P_{1,3}=x_0=0$ as we have chosen here, then $l^{\alpha}$ is completely irrelevant in the degree $(1,4)$ polynomials, being multiplied as it is by zero. Its correlation to another polynomial is therefore irrelevant. Likewise, on the resolution, the second term in (\ref{p5split}) is zero in the coordinate ring of the Calabi-Yau 3-fold. As such its correlation to the degree $(1,4)$ polynomials simply does not matter in terms of describing general vacua. The second point we should raise is that  polynomial long division, and thus the decomposition (\ref{p5split}), is not unique. Thus there are many ways in which we can split up a quintic in the form  (\ref{p5split}) and therefore ways in which that polynomial can be correlated with the $(1,4)$ polynomials. For the purposes pursued in matching complex moduli and showing that Yukawa couplings agree across the duality, any split of the form (\ref{p5split}) will work. The actual physical system should pick out such a split uniquely, however. We believe that it is the mapping to the extra K\"ahler modulus that appears on the resolution side of the duality relative to the deformation that fixes this ambiguity. Indeed, note that the split (\ref{p5split}) is separating the polynomial in to terms which are of nodal form and everything else. That is, into pieces corresponding to active and spectator complex structure degrees of freedom in our earlier nomenclature. Although a full detailing of this part of the field space map is beyond the remit of this paper, we will return to this subject shortly to offer more of an explanation.

\vspace{0.2cm}

Our claim, then, is that a good map between the degrees of freedom of the two theories is given by correlating the defining data in the manner described around (\ref{14s}) and (\ref{p5split}) and then using the same data to define the two dual compactifications. If this is correct, then considering a linear fluctuation of the defining data should allow the formation of a map between the infinitessimal moduli of the two theories, which can then be checked for consistency with the computation of the tangent to the moduli space on the two sides of the duality. We begin an analysis of this verification by detailing the computation of the moduli of the theories on the two sides of the duality, describing infinitesimal fluctuations around a given background.

\subsection{Moduli of the deformation}

A description of the moduli of the Calabi-Yau manifold itself can be obtained from the adjunction and Euler sequences associated to $X_D$ in (\ref{one}). Referring to the ambient space as $A$ we have,
\begin{eqnarray} 
     0 \to TX_D \to TA|_{X_D} \to {\cal O}(1,3) \oplus {\cal O}_{X_D} (0,5)\to 0 \;,
\end{eqnarray}
and
\begin{eqnarray} 
     0 &\to& {\cal O}_{X_D}^2 \to {\cal O}_{X_D}(0,1)^{\oplus 5} \oplus {\cal O}_{X_D}(1,3)\oplus {\cal O}_{X_D}(1,0) \to TA|_{X_D} \to 0\;.
\end{eqnarray}
Analysing the long exact sequences in cohomology associated to these short exact sequences one finds the following description of $H^1(TX_D)$ and $H^1(TX_D^{\vee})$ in terms of polynomial equivalence classes.
\begin{eqnarray}\label{quintfluct}
&H^1(TX_D)&:\left\{ \begin{array}{c} \delta P_{0,5} \sim \delta P_{0,5} + h P_{0,5}+l^i_{0,1}
 \partial_{y^i} {P}_{0,5}\\  
 \delta P_{1,3} \sim \delta P_{1,3} + m P_{1,3} +l^i_{0,1} \partial_{y^i} {P}_{1,3}+ l_{1,3} \partial_{x^0} {P}_{1,3}+ l_{1,0} \partial_{x^1} {P}_{1,3}\end{array} \right. \\  \nonumber 
 &H^1(TX_D^{\vee})&: \mathbb{C}\end{eqnarray}

Here  $l_{0,1}^i$, $l_{1,3}$ and $l_{1,0}$ are arbitrary polynomials of the multi-degree indicated by their subscript and $h$ and $m$ are arbitrary constants.

The quintic equivalence classes above are easy to understand. These are just the deformations of the defining quintic, where two deformations are equivalent if they could be related by an infinitesimal coordinate transformation $y_i \to y_i + l^i_{0,1}$ where the $y_i$ are the five coordinates of weight $(0,1)$. The remaining equivalence involving $h$ is in fact redundant with these coordinate changes and simply corresponds to the fact that scaling the defining relation does not change the Calabi-Yau locus. We choose to keep terms such as this in our analysis for reasons that will become clear as we proceed. Similarly the $\delta P_{1,3}$ fluctuations are the deformations of the defining relation of that degree up to coordinate transformations of the $y_i$ and the $x_a$. Note that any $P_{1,3}$ fluctuation here is in fact equivalent to zero by the 4th term on the right of the equivalence relation. This is the manifestation in this fluctuation analysis of the fact that one can use coordinate redefinitions to set the defining relation $P_{1,3}$ to a specific polynomial, say $x_0$. We will continue to make the choice $P_{1,3}=x_0$ in this subsection.

For the brane moduli we would conventionally fix the base manifold entirely and look at fluctuations of the brane degrees of freedom. These are then given by $H^0({\cal N}_{{\cal C}_D}|_{{\cal C}_D})$ a description of which can be obtained from the following Koszul sequence (which we have broken into two short exact pieces by the introduction of a kernel/cokernel $k_1$).
\begin{eqnarray}
0 \to {\cal O}_{X_D}(-1,-4) \to {\cal O}_{X_D}^{\oplus 2} \to k_1 \to 0\\ \nonumber
0 \to k_1 \to {\cal O}_{X_D}(1,4) \to {\cal O}_{{\cal C}_D}(1,4) \to 0
\end{eqnarray}
Using the associated long exact sequences in cohomology, we arrive at the following description of the tangent to the brane position moduli space.
\begin{eqnarray} \label{initial}
\delta P_{1,4}^{\alpha} \sim \delta P_{1,4}^{\alpha} + A^{\alpha}_{\beta} P^{ \beta}_{1,4} + L_{0,1}^{\alpha} P_{1,3}
\end{eqnarray}
In this expression $\alpha=1,2$, $A$ is a matrix of arbitrary constants and the $L_{0,1}^\alpha$ are two arbitrary degree $(0,1)$ polynomials. The equivalence of perturbations here just corresponds to changes in the defining $(1,4)$ polynomials that would not change the locus of the brane stack on the Calabi-Yau 3-fold.

However, here we need to be more careful. We do not wish to consider fluctuations of the brane moduli wherein the base manifold is fixed, but rather fluctuations of the complete system. Given the fact that the $l_{0,1}^i$, $l_{1,3}$ and $l_{1,0}$ appearing in (\ref{quintfluct}) arise from coordinate transformations, and that these coordinate transformations would affect the description of the locus wrapped by the brane too, we need to modify (\ref{initial}) as follows.
\begin{eqnarray} \label{quintbranefluct}
H^0({\cal N}_{{\cal C}_D}|_{{\cal C}_D}): \delta P_{1,4}^{\alpha} &\sim &\delta P_{1,4}^{\alpha} + A^{\alpha}_{\beta} P^{ \beta}_{1,4} + L_{0,1}^{\alpha} P_{1,3} \\ \nonumber &&+l^i_{0,1} \partial_{y^i} P^{ \alpha}_{1,4} + l_{1,3}\partial_{x^0}P^{ \alpha}_{1,4}+l_{1,0}\partial_{x^1}P^{\alpha}_{1,4}
\end{eqnarray}
Note that, because of their origin in terms of coordinate transformations that affect the whole system, the $l^i_{0,1}$, $l_{1,3}$ and $l_{1,0}$ appearing here are the same as those in (\ref{quintfluct}). Thus, while the total dimension of the moduli space is as usual, in this description one can assign moduli to different sources, depending on how the coordinate transformations are used. For example one could either use redefinitions of $x_0$ (as embodied in $l_{1,3}$) to trivialize fluctuations of $P_{1,3}$ or to trivialize some fluctuations of the brane stack. In other words, one could either use the coordinate definition freedom to fix the $(1,3)$ defining relation or to constrain the form of the defining equations of the brane locus.

In the end, the infinitesimal singlet fluctuations of the system are described by the equivalence classes (\ref{quintfluct}) and (\ref{quintbranefluct}). And it is these that we should compare to the moduli of the resolution manifold, in light of our proposal for the map of degrees of freedom between the dual systems as discussed earlier in Section \ref{branemap}.

\subsection{Moduli of the resolution}

A similar analysis can be given for the resolution side of the duality. In particular we have the adjunction sequence,
\begin{eqnarray}
     0 \to TX_R \to TA|_{X_R} \to {\cal O}(1,4)^{\oplus 2} \to 0 \;,
\end{eqnarray}
and the Euler sequence,
\begin{eqnarray}
     0 \to {\cal O}_{X_R}^2 \to {\cal O}_{X_R}(0,1)^{\oplus 5} \oplus {\cal O}_{X_R}(1,3)\oplus {\cal O}_{X_R}(1,0) \to TA|_{X_R} \to 0\;,
\end{eqnarray}
in this case. These lead to the following description of the moduli of the Calabi-Yau manifold.
\begin{eqnarray} \label{rescs}
&H^1(TX_R)&: \delta P_{1,4}^{\alpha}  \sim \delta P_{1,4}^{\alpha} + A^{\alpha}_{\beta} P_{1,4}^{ \beta} \\ \nonumber &&+l^i_{0,1} \partial_{y^i} P^{ \alpha}_{1,4} + l_{1,3}\partial_{x^0}P^{ \alpha}_{1,4}+l_{1,0}\partial_{x^1}P^{\alpha}_{1,4}
\\ \nonumber
&H^1(TX_R^{\vee})&: \mathbb{C}^2
\end{eqnarray}
In the above, $A$ is a matrix of constants and the $l_{0,1}^i$, $l_{1,3}$ and $l_{1,0}$ are arbitrary polynomials of the indicated degree corresponding to ambient space coordinate transformations as was described in our analysis of the quintic fluctuations. Note that two of the constants in $A$ here are redundant with the coordinate transformations if we restrict our attention to purely these fluctuations, however we choose to keep them here for simplicity.

Additionally, Koszul sequences for the line bundles appearing in the normal bundle of the  brane stack given in (\ref{two}) can be split up as follows.
\begin{eqnarray}
0 \to {\cal O}_{X_R}(0,-5) \to {\cal O}_{X_R}\oplus{\cal O}_{X_R}(1,-2) \to k_2 \to 0\\ \nonumber
0 \to k_2 \to {\cal O}_{X_R}(1,3) \to {\cal O}_{{\cal C}_R}(1,3) \to 0
\end{eqnarray}
\begin{eqnarray}
0 \to {\cal O}_{X_R}(-1,-3) \to {\cal O}_{X_R}\oplus {\cal O}_{X_R}(-1,2) \to k_3 \to 0\\ \nonumber
0 \to k_3 \to {\cal O}_{X_R}(0,5) \to {\cal O}_{{\cal C}_R}(0,5) \to 0
\end{eqnarray}
These give rise to the following description of the brane moduli.
\begin{eqnarray} \label{branefluctres}
H^0({\cal N}_{{\cal C}_R}|_{{\cal C}_R}): \left\{ \begin{array}{c} \delta P_{0,5} \sim \delta P_{0,5}+ h P_{0,5}^{} + B \epsilon_{\alpha \beta} l^{\alpha} q^{\beta}+L^{\alpha}_{0,1}q^{ \alpha}+ l^i_{0,1}
 \partial_{y^i} {P}_{0,5}\\ 
 \delta P_{1,3} \sim \delta P_{1,3} + m P_{1,3}^{}+ l^i_{0,1} \partial_{y^i} {P}^{}_{1,3}+ l_{1,3} \partial_{x^0} {P}_{1,3}+ l_{1,0} \partial_{x^1} {P}_{1,3}\end{array} \right. 
\end{eqnarray}
In this expression the $L^{\alpha}_{0,1}$ are two arbitrary linears in the $y$'s and the quantities $h$ and $B$ are arbitrary constants.

Note that we have once again included additional terms in (\ref{branefluctres}) relative to the analysis over a fixed base, as we did on the deformation side of the duality. One can use this freedom to define coordinates to choose $P_{1,3} =x_0$ in this setting too. Thus our simplifying choice in this regard is consistently achievable on both sides of the duality.

In the end, the infinitesimal singlet fluctuations of the resolution are described by the equivalence classes (\ref{rescs}) and (\ref{branefluctres}). We now go on to compare these to the deformation degrees of freedom in light of the proposed moduli map discussed earlier in Section \ref{branemap}

\subsection{Comparing the moduli} \label{comp1}

Comparing the last two subsections, we see that the descriptions of the singlet moduli of the deformation and resolution sides of the duality only differ by a few terms. Here we will write out the equivalence classes describing the physical infinitesimal fluctuations of the defining polynomials in the two cases, only keeping those terms in the equivalence classes that differ on the two sides of the duality.
\begin{eqnarray} \label{dif}
\begin{array}{c|c} 
\text{Deformation:} & \text{Resolution:} \\ \hline
\delta P_{0,5} \sim \delta P_{0,5} & \delta P_{0,5} \sim \delta P_{0,5} +  B \epsilon_{\alpha \beta} l^{\alpha} q^{\beta} + L_{0,1}^{\alpha} q^{\alpha} \\
\delta P^{\alpha}_{1,4} \sim \delta P^{\alpha}_{1,4} + L_{0,1}^{\alpha} x_0 & \delta P_{1,4}^{\alpha} \sim \delta P_{1,4}^{\alpha}
\end{array}
\end{eqnarray}
On both sides we will use coordinate freedom to set $P_{1,3}=x_0$ and thus no fluctuations are considered of this degree.

In isolation, the discrepancies in the above two descriptions make it appear as though the moduli are described differently in terms of polynomial fluctuations on the two sides of the duality. However, when combined with our restricted form of the defining polynomials as embodied in (\ref{14s}) and (\ref{p5split}), the situation becomes clearer. Consider a change of the $(1,4)$ defining relations of the form $\delta P_{1,4}^{\alpha}= L_{0,1}^{\alpha} x_0$. Such a change is a physical modulus on the resolution side of the duality but not on the deformation according to (\ref{dif}). According to (\ref{14s}) and (\ref{p5split}), however, such a change in $P_{1,4}$ is correlated with a change $\delta P_{0,5} = -\epsilon_{\alpha \beta} L_{0,1}^{\alpha} q^{\beta}$. This change in the $(0,5)$ polynomial is a physical modulus on the deformation side of the transition but not on the resolution. So the correlation indeed correctly links the polynomial structure such that physical deformations always map to physical deformations across the duality. The one seeming exception to this statement is a fluctuation of the $(1,4)$ polynomials of the form $\delta P_{1,4}^{\alpha} = \eta P_{1,4}^{\alpha}$. A change of this form is correlated in our ansatz to a change of the $(0,5)$ polynomial of the form $\delta P_{0,5} = 2 \eta \epsilon_{\alpha \beta} l^{\alpha} q^{\beta}$. Together these changes do not correspond to a physical modulus on the resolution side, as scaling the $P_{1,4}^{\alpha}$ defining polynomials does not change the geometry and the change in $P_{0,5}$ is removed from the description of the moduli by the term proportional to $B$ in (\ref{dif}). On the deformation side however, this one complex parameter family of perturbations does correspond to a physical change of the quintic defining relation. We conjecture that it is this extra degree of freedom on the deformation side that maps, in a non-trivial fashion, to the addition K\"ahler modulus, $T_R^0$, found in the resolution geometry.

The mapping to the new K\"ahler modulus proposed in the last paragraph obeys some consistency checks. In particular, our proposal is that the relationship between the resolution side K\"ahler modulus and $\eta$ is inverse in some way, with large values of $\eta$ corresponding to small K\"ahler moduli (although we do not give the specific form). Note then, that when $\eta$ becomes very large one can, by using the homogeneous scaling, think of this as taking $P_{0,5}'$ in (\ref{p5split}) to be very small. On the deformation side of the transition we thus approach the nodal variety in such a limit, which is precisely what is achieved by taking the new K\"ahler modulus to zero on the resolution side of the duality. 

One can consider other singular limits to perform consistency checks of the duality map on moduli space that we have proposed. For example, one could consider taking $l^{\alpha}=0$ and everything else general. This is clearly a singular point in moduli space on the resolution side of the duality with the defining relations of $X_R$ now factorizing. The corresponding singular structure on the deformation side of the duality is not so clear. There, setting $l^{\alpha}=0$ does not make the $(1,4)$ equations defining the brane singular because $x_0=0$ in the coordinate ring, rendering the relevant terms in the $P_{1,4}^{\alpha}$ unphysical anyway. The quintic defining relation, however, becomes purely $P_{0,5}'$ in such a limit, that is the remainder of a long division of a general quintic by $q^{\alpha}$. One can show, by direct computation, that such a quintic is indeed singular for specific choices of the ambiguity in defining that long division, as described at the start of Section \ref{branemap}. One can use, for example, a Gr\"obner basis with a specific choice of ordering to obtain this result. Thus this is consistent with the possibility that the singular behavior does indeed match across the correspondence. 

The real check of the proposed mapping of moduli spaces is, however, given by showing that elements of the four dimensional effective field theories agree when such an identification is employed. In particular, in this paper we will show that the mapping described in this section leads to Yukawa couplings matching precisely as holomorphic functions of the moduli across the duality. To provide this evidence we require a compactification involving manifolds dressed with bundles rather than branes, such that the desired Yukawa couplings exist and are computable. It is to the construction of such an example that we now turn.

\section{The moduli map in a bundle example} \label{mrbun}

Let us consider an example based around bundles that is very close to the case with branes just analyzed. On the deformation side we choose the following manifold,
\begin{eqnarray}
X_{D}&=&\left[\begin{array}{ccccccc|cc} 0&0&0&0&0&1&1&1&0\\1&1&1&1&1&3&0&3&5 \end{array} \right] \;,
\end{eqnarray}
and bundle,
\begin{eqnarray} \label{mond}
0 &\to& V_D \to {\cal O}(0,1)^5 \to {\cal O}(1,5) \to 0 \;.
\end{eqnarray}
On the resolution side of the duality we then have the following manifold,
\begin{eqnarray}
X_R&=&\left[\begin{array}{ccccccc|cc} 0&0&0&0&0&1&1&1&1\\1&1&1&1&1&3&0&4&4 \end{array} \right] ,
\end{eqnarray}
and bundle,
\begin{eqnarray}
0 &\to& V_R \to {\cal O}(0,1)^3 \oplus {\cal O}(0,2) \oplus {\cal O}(1,0) \to {\cal O}(1,5) \to 0.
\end{eqnarray}

In the setup of these two vacua, we see that they are once again defined in terms of the same set of polynomial data as functions of the same ambient coordinates. In particular, the deformation geometry is defined by degree $(1,3)$ and $(0,5)$ polynomials specifying the manifold and five degree $(1,4)$ polynomials specifying the bundle (that is the maps in the monad sequence (\ref{mond})). On the resolution side of the transition we have two degree $(1,4)$ polynomials describing the Calabi-Yau, and three more and a $(1,3)$ and a $(0,5)$ polynomial describing the bundle $V_R$.

For reasons analogous to those seen in Section \ref{branemap}, the map on moduli space cannot quite be determined simply by taking the same defining polynomials on the two sides of the duality. However, we claim that the moduli map here is essentially unchanged from the brane based case just discussed. On the deformation side of the duality pick two of the degree $(1,4)$ polynomials describing the bundle map and denote them by $P_{1,4}^{\alpha}$. Denote the remaining degree $(1,4)$ polynomials by $P_{1,4}^i$. Then our claim is that the moduli map in this case is precisely the same as that given at the start of Section \ref{branemap}, with the addition that the three polynomials $P_{1,4}^i$ which appear in the bundle on both sides of the conifold transition are identified on the two sides of the duality.

To provide some initial evidence that this moduli mapping is correct, before proceeding to show that it induces a matching of the Yukawa couplings of the theories, we will now show that it is consistent with the description of the tangent to the moduli spaces of these compactifications.

\subsection{Moduli of the deformation}

Since the deformation manifold is unchanged from our earlier example, the complex structure and K\"ahler moduli are still given by (\ref{quintfluct}), which we reproduce here.  
\begin{eqnarray}\label{quintfluct2}
&H^1(TX_D)&:\left\{ \begin{array}{c} \delta P_{0,5} \sim \delta P_{0,5} + h P_{0,5}\\  
 \delta P_{1,3} \sim \delta P_{1,3} + m P_{1,3} \end{array} \right. \\  \nonumber 
 &H^1(TX_D^{\vee})&: \mathbb{C}
 \end{eqnarray}
We will not write out the full sequence chasing for the bundle moduli, but the results are as follows.
\begin{eqnarray} \label{vvdef}
H^1( V_D \otimes V^{\vee}_D): \delta P_{1,4}^A \sim \delta P_{1,4}^A + M^A_B P_{1,4}^{B} + L_{0,1}^A P^{}_{1,3}
\end{eqnarray}
Here $A=1,\ldots, 5$, the $M^A_B$ are arbitrary constants and the $L_{0,1}^A$ are arbitrary polynomials of the indicated degree. In both of the above equations we are omitting terms defining the equivalence classes that are associated to infinitesimal coordinate transformations in the interests of brevity.

\subsection{Moduli of the resolution}

Since the resolution manifold is also unchanged from our earlier brane based example, the complex structure and K\"ahler moduli are still given by (\ref{rescs}), which we again reproduce.
\begin{eqnarray} \label{rescs2}
&H^1(TX_R)&: \delta P_{1,4}^{\alpha}  \sim \delta P_{1,4}^{\alpha} + A^{\alpha}_{\beta} P_{1,4}^{ \beta} 
\\ \nonumber
&H^1(TX_R^{\vee})&: \mathbb{C}^2
\end{eqnarray}
In this case, a standard computation reveals that the bundle moduli are associated to the following equivalence classes.
\begin{eqnarray} \label{vvres}
H^1( V_R \otimes V^{\vee}_R): \left\{ \begin{array}{c} \delta P_{0,5}\sim  \delta P_{0,5} + h P_{0,5}^{}  +B \epsilon_{\alpha \beta} l^{\alpha} q^{\beta} + c_{\alpha i}(l^{\alpha}q^{i} - q^{\alpha} l^{i}) + L^{\alpha}_{0,1}(c q^{ \alpha}- C_{0,3} l^{\alpha}) \\ \delta P_{1,3}\sim \delta P_{1,3} + m P_{1,3}\\\delta P_{1,4}^i \sim \delta P_{1,4}^i + M^i_A P_{1,4}^{A} + L_{0,1}^i P_{1,3}\end{array} \right.\;,
\end{eqnarray}
where the $c_{\alpha i}$ are arbitrary constants and we define,
\begin{eqnarray}
P_{1,4}^{ i} = l^{i} x_0 + q^{i} x_1 \;,
\end{eqnarray}
for $i=1,\ldots, 3$. We also define
\begin{eqnarray}
P_{1,3} = c x_0 + C_{0,3} x_1 \;.
\end{eqnarray}
In what follows we will use coordinate freedom to set $P_{1,3}=x_0$ such that $c=1$ and $C_{0,3}=0$.
Note that the index $A$ in the above expressions runs over both the monad map and defining equation degree $(1,4)$ polynomials. In all of the above expressions, we are once again omitting terms associated to coordinate transformations in the interests of brevity.

\subsection{Comparing the moduli}

Comparing the moduli of the deformation and resolution side of the duality as presented in the last two subsections, we see that once again the descriptions of the singlet moduli differ by only a few terms. Here we write out the equivalence classes describing the physical infinitesimal fluctuations of the defining polynomials in the two cases, keeping only those terms in the equivalence classes that differ on the two sides of the duality.
\begin{eqnarray} \label{dif2}
\begin{array}{c|c} 
\text{Deformation:} & \text{Resolution:} \\ \hline
\delta P_{0,5} \sim \delta P_{0,5} & \delta P_{0,5} \sim \delta P_{0,5} +  B \epsilon_{\alpha \beta} l^{\alpha} q^{\beta}  + c_{\alpha i}(l^{\alpha}q^{i} - q^{\alpha} l^{i})+ L_{0,1}^{\alpha} q^{\alpha} \\
\delta P^{\alpha}_{1,4} \sim \delta P^{\alpha}_{1,4} + M^{\alpha}_i P_{1,4}^i+ L_{0,1}^{\alpha} x_0 & \delta P_{1,4}^{\alpha} \sim \delta P_{1,4}^{\alpha}  \\ 
\delta P_{1,4}^i \sim \delta P_{1,4}^i&\delta P_{1,4}^i \sim \delta P_{1,4}^i
\end{array}
\end{eqnarray}
The way in which most of the differences exhibited in (\ref{dif2}) are accounted for by our moduli map given by (\ref{14s}) and (\ref{p5split}) is described in Section \ref{comp1}. The only differences left to understand are those parameterized by $M_i^{\alpha}$ on the deformation side and $c_{\alpha i}$ in the resolution geometry.

Consider a fluctuation $\delta P_{1,4}^{\alpha} = M_i^{\alpha} P_{1,4}^i$. Such a change corresponds to a physical deformation on the resolution side of the duality but not on the deformation. Note that such a fluctuation corresponds to changes $\delta l^{\alpha} = M_i^{\alpha} l^i$ and $\delta q^{\alpha} = M_i^{\alpha}q^i$. According to our procedure, such a fluctuation is correlated, via (\ref{14s}) and (\ref{p5split}), to another variation $\delta P_{0,5} =\epsilon_{\alpha \beta} M^{\alpha}_i l^i q^{\beta} + \epsilon_{\alpha \beta} l^{\alpha} M_i^{\beta} q^i = \epsilon_{\alpha \beta} M^{\alpha}_i (l^i q^{\beta}-q^i l^{\beta})$. Here we have ignore second order terms in the fluctuations in order to make contact with the linearized moduli computation. We see therefore, that the fluctuation $\delta P^{\alpha}_{1,4} = M^{\alpha}_i P^i_{1,4}$ is correlated with a fluctuation in the $(0,5)$ polynomial which, thanks to the term proportional to $c_{\alpha i}$ in (\ref{dif2}) is physical in the deformation but not in the resolution. We therefore once again have a perfect match of physical fluctuations across the duality, once the existence of the extra K\"ahler modulus on the resolution is taken into account.

\section{Testing the moduli map -- Comparison of holomorphic Yukawa couplings across the duality} \label{yukkysection}
We would like to do better than simply looking at tangents to moduli spaces, and explicitly compute some functions appearing in the four-dimensional Lagrangians, demonstrating that they match in related theories. Some of the easiest such functions to compute in heterotic theories are the superpotential Yukawa couplings. These are complicated holomorphic functions of the complex structure and bundle moduli in general and, as such, have the potential to provide strong evidence that the putative duality indeed relates equivalent theories.

In order to keep the discussion as transparent as possible, we will work in terms of examples. In what follows we will see both $E_6$ and $SU(5)$ examples with non-vanishing couplings that match across the duality. The couplings in an $E_6$ model are of the form ${\bf 27}^3$ and in an $SU(5)$ model we have both ${\bf 10} \;{\bf 10} \; {\bf 5}$ and $\overline{\bf 5}\;\overline{\bf 5}\;{\bf 10}$ Yukawas present. In both cases, the couplings are generically complicated holomorphic functions of both the complex structure and bundle moduli. Therefore, matching them requires at least a partial knowledge of the mapping between both moduli and matter fields across the duality. We will discuss this further in what follows. 

We will only require explicit form for the Yukawa couplings in the $E_6$ case in the examples we will present. There is a natural mapping from three copies of $H^1(V)$, the cohomology that describes the generations, to $H^3(\wedge^3 V) = H^3({\cal O}) =\mathbb{C}$ in the case of an $SU(3)$ bundle. It is the image for any choice of three elements of $H^1(V)$ that is the superpotential Yukawa coupling between the associated matter fields \cite{Distler:1987ee,Anderson:2010vdj}. 

\subsection{A first example}

For some examples, it is easy to give an explicit matching of Yukawa couplings across the duality. For the case given in Section \ref{mrbun}, which was chosen so that the field space mapping could be illustrated in the most concise manner possible, the Yukawa couplings do match. However, this matching is trivial as those compactifications correspond to $SO(10)$ theories with no matter fields in the ${\bf 10}$ representation, and hence all the Yukawa couplings vanish. Here we wish to study a less trivial example.

Take the following case as an illustration. On the deformation side we have
\begin{eqnarray}
X_D=\left[ \begin{array}{c|c} \mathbb{P}^4 &5 \end{array}\right]\;,
\end{eqnarray}
with bundle
\begin{eqnarray} \label{everywhereVd}
0 \to V_D \to {\cal O}(2) \oplus {\cal O}(1)^{\oplus 3} \to {\cal O}(5) \to 0\;.
\end{eqnarray}
On the resolution side we have
\begin{eqnarray} \label{toricintersec}
X_R=\left[\begin{array}{ccccccc|cc} y_0&y_1&y_2&y_3&y_4&x_0&x_1&p^1_{(1,4)}&p^2_{(1,4)}\\0&0&0&0&0&1&1&1&1 \\ 1&1&1&1&1&3&0&4&4\end{array} \right]\;.
\end{eqnarray}
As before, this notation describes a manifold which is a complete intersection in a toric variety. The left hand portion of the matrix describes the behavior of ambient space coordinates under two scalings and the right hand portion gives the degrees of the two defining relations of the variety relative to these scalings. The bundle over this manifold is found via the duality construction to be the following:
\begin{eqnarray}
0 \to V_R \to {\cal O}(0,2)^{\oplus 2} \oplus {\cal O}(0,1) \oplus {\cal O}(1,0) \to {\cal O}(1,5) \to 0\;.
\end{eqnarray}

The generations of $V_D$ are described by equivalence classes of polynomials:
\begin{eqnarray} \label{p5}
P_5 \sim P_5 + \alpha p_{(5)} + \sum_{a=1}^3 \beta_a m_{(4)}^a + \gamma m_{(3)}\;.
\end{eqnarray}
Here, $p_{(5)}$ is the defining relation of the quintic and the $m_{(d)}$ are degree $d$ maps associated to the defining sequence of $V_D$ (\ref{everywhereVd}).

A Yukawa coupling on the deformation side of the duality is obtained by taking three representatives of the equivalence classes in (\ref{p5}) associated to the three fields of interest and multiplying them together. One then regards this as a representative $r_{15}$ of the following equivalence class, which describes elements of $H^3(\wedge^3 V_D):$
\begin{eqnarray} \label{p15}
P_{15} \sim P_{15} + A p_{(5)} + \sum_{a=1}^3 B_a m_{(4)}^a+ D m_{(3)}.
\end{eqnarray}
Using the equivalence relation above one can change $r_{15}$ to be a multiple of some, arbitrarily chosen, representative of the class (which is one dimensional) with the same choice being used to compute all couplings. The multiple that one finds in this manner is then the Yukawa coupling.

The generations of $V_R$ are described by the following equivalence classes of polynomials:
\begin{eqnarray}
P_{1,5} \sim P_{1,5} + \alpha m_{(0,5)} +\sum_{a=1}^2 {\beta}_a p^a_{(1,4)} + {\beta}_3 m_{(1,4)}  + \sum_{i=1}^2 {\gamma}_i m^i_.
\end{eqnarray}
Although these equivalence classes naively look very different to (\ref{p5}) one can spot a clear isomorphism as follows. By use of a change of coordinates, one of the monad maps $m_{(1,3)}$ can be taken to simply be $x_0$ from (\ref{toricintersec}) without loss of generality. This then means that any polynomial proportional to $x_0$ is equivalent to zero, and thus the equivalence classes can be encoded in the following manner:
\begin{eqnarray} \label{p05}
x_1 P_{0,5} \sim x_1 P_{0,5} + x_1 \tilde{\alpha}m_{(0,5)}+x_1\sum_{a=1}^2 {\beta}_a p^a_{(0,4)} + x_1{\beta}_3 m_{(0,4)}   + x_1{\gamma}_2 m^2_{(0,3)}.
\end{eqnarray}
Here we chose $m_{1,3}^1=x_0$ and the associated term in the equivalence class description has been removed as we have used it to remove the $x_0$ dependence as described above. There is now a clear one-to-one correspondence of equivalence classes between (\ref{p05}) and (\ref{p5}) with the classes in the former just being classes in the latter multiplied by $x_1$. To make this identification we must set $m_{(0,5)}=p_{(5)}$, $p_{(0,4)}^a= m_{(4)}^a$ for $a=1,2$, $m_{(0,4)}=m^3_{(4)}$ and $m_{(0,3)}^2 =m_{(3)}$, determining defining polynomials of the bundle and manifold on the resolution side in terms of similar objects appearing in the deformation compactification. This type of identification matches exactly with the proposed duality map from earlier sections.

\vspace{0.1cm}

The described mapping of family equivalence classes is a proposal for the charged part of the field mapping between the two theories across the duality. While, given the above structure, the proposed field matching seems natural, the first non-trivial test that this is correct comes when we use it to compare Yukawa couplings on the deformation and resolution compactifications. The polynomial equivalence class associated to the one-dimensional cohomology $H^3(\wedge^3 V_R)$ is given by the following:
\begin{eqnarray}
P_{3,15} \sim P_{3,15} + {A}m_{(0,5)} +\sum_{a=1}^2 {B}_a p^a_{(1,4)}  + {B}_3 m_{(1,4)} + \sum_{i=1}^2 {D}_i m^i_{(1,3)}.
\end{eqnarray}
Making, as we must, the same choice for $m_{1,3}^1$ as we did above, this equivalence class can be represented as follows:
\begin{eqnarray}\label{p015}
x_1^3 P_{0,15} \sim x_1^3  P_{0,15} + x_1^3  \tilde{A}m_{(0,5)}  +x_1^3 \sum_{a=1}^2 \tilde{{B}_a} p^a_{(0,4)} + x_1^3 \tilde{{B}_3} m_{(0,4)} + x_1^3  \tilde{{D}_2} m^2_{(0,3)}.
\end{eqnarray}
One can then use the family representatives (\ref{p05}) and the class (\ref{p015}) to obtain the Yukawa couplings in the same manner as we did on the deformation side above.

Comparing (\ref{p05}) and (\ref{p015}) to (\ref{p5}) and (\ref{p15}) the relationship between the Yukawa coupling computations on the two sides of the duality now becomes clear. The computation being performed on the resolution side is identical to that being performed for the deformation geometry, except for some overall factors of $x_1$ which spectate through the process and do not affect the result. As such, with the field mapping we have described here, the complete set of Yukawa couplings in these two compactifications agree as functions of the complex structure and bundle moduli across the entire moduli space of the compactifications. This constitutes considerable evidence for duality beyond the simple matching of integers that results from showing that the spectrum agrees between the two theories. In the example analyzed in this section there are 95 families. As such, the above results demonstrate that 147,440 independent couplings agree, as holomorphic functions of the moduli fields, across the two sides of the putative duality.

\vspace{0.2cm}

An important question to ask is whether the ambiguity in the structure of the moduli space map seen in Sections \ref{branemap}  and \ref{mrbun} affects whether or not we achieve matching of the Yukawa couplings in such examples. We can see that this is not the case. Nothing special had to be assumed about the structure of the defining polynomials above. Furthermore, the ambiguity in determining $P_{0,5}'$ does not affect the Yukawa couplings according to the above computations. This is simply because this ambiguity would correspond to terms which are proportional to $q^{\alpha}$ and such terms are quotiented out in the equivalence classes that play a key role in this section. The structure of these computations also give further credence to the suggestion that the nodal deformation of the quintic maps to the extra K\"ahler modulus on the resolution side of the duality. This is because this deformation drops out of the Yukawa coupling computation for the same reasons. This is required if the proposal is to be true, because the Yukawa coupling on the resolution side cannot, of course, depend upon a K\"ahler modulus.

\vspace{0.2cm}

It is easy to provide an analysis of this type, holding everywhere in moduli space, for every case where the conifold is described as a $\mathbb{P}^1$ split, where the bundles are described as two term monads, and where the rank of the last terms in the two term monads is one. In such cases, in the associated GLSM (0,2) target space duality, the so called ``target space duality tuning" required to make the GLSMs equivalent in a non-geometric phase holds everywhere in moduli space. One could be concerned that such matching of Yukawa couplings is merely an artifact of this structure and the fact that the couplings and numbers of generations do not depend upon the K\"ahler moduli. In the next example, we present a case where the Yukawa couplings match everywhere in moduli space, even when the tuning of moduli used in target space duality does not hold.

\subsection{A second example}

We can provide other examples where the Yukawa couplings of the two dual theories can be shown to match everywhere in moduli space, but where the specialization satisfied by the case in the proceeding subsection does not hold. In general we don't know how to explicitly write down the matter field identification necessary to pursue such an analysis in, for example, cases described by general two term monad bundles over CICYs related by $\mathbb{P}^1$ splits. However, the following dual pair, also considered in \cite{Anderson:2016byt}, is an example which exhibits all of the same structure as general constructions of that form, but where we can show the Yukawa couplings match across the duality for other reasons.

For the deformation side we take the manifold to be given by
\begin{eqnarray} \label{same1}
X_D=\left[ \begin{array}{c|c} \mathbb{P}^1& 2 \\ \mathbb{P}^3 &4 \end{array} \right]\;,
\end{eqnarray}
and the bundle by
\begin{eqnarray} \label{vd1}
0\to V_D \to {\cal O}(1,0)^2 \oplus {\cal O}(0,1)^4 \to {\cal O}(2,4)\to 0  \;.
\end{eqnarray}
For the resolution side of the transition we take the manifold to be given by
\begin{eqnarray} \label{same2}
X_R=\left[ \begin{array}{c|cc} \mathbb{P}^1& 1&1\\ \mathbb{P}^1& 1&1\\ \mathbb{P}^3 &4&0 \end{array} \right] \;,
\end{eqnarray}
and then find that the bundle is given by the following:
\begin{eqnarray}
0 \to V_R \to {\cal O}(1,0,0)\oplus {\cal O}(0,1,0) \oplus {\cal O}(0,0,1)^4 \oplus {\cal O}(0,2,4)\\ \nonumber \to {\cal O}(0,1,4) \oplus {\cal O}(1,2,4) \to0 .
\end{eqnarray}
This example has $84$ generations and a single $5-\overline{5}$ pair of Higgs multiplets. 

The first thing to note is that the transition from (\ref{same1}) to (\ref{same2}) is what is termed an ``ineffective split" \cite{Candelas:1987kf}. The common point in moduli space of the two varieties is infact smooth, as the singular locus would be given by the following variety
\begin{eqnarray}
\left[\begin{array}{c|cccc} \mathbb{P}^1 & 1&1&1&1\\ \mathbb{P}^3 &4&4&0&0\end{array}\right] = \varnothing\;,
\end{eqnarray}
which is the empty set as there is no solution to two linears in $\mathbb{P}^1$. As such, (\ref{same1}) and (\ref{same2}) are isomorphic geometries. Nevertheless, one might think that the two bundles being considered over these varieties are different. By examining cohomology one can derive the following isomorphisms of line bundles over the two different descriptions of the variety
\begin{eqnarray}
{\cal O}_{X_R}(a,b,c) \equiv {\cal O}_{X_D}(a+b,c) .
\end{eqnarray}
As such, writing $V_R$ in a form which is compatible with the description of the Calabi-Yau 3-fold given in (\ref{same1}) we arrive at the following:
\begin{eqnarray} \label{VR2}
0 \to V_R \to {\cal O}(1,0)^2 \oplus {\cal O}(0,1)^4 \oplus {\cal O}(2,4) \to {\cal O}(1,4) \oplus {\cal O}(3,4) \to 0 .
\end{eqnarray}
Although it is not obvious that $V_R$ as given in (\ref{VR2}) and $V_D$ as given in (\ref{vd1}) are isomorphic, they are. This can be proven using the following theorem.

\vspace{0.2cm}

{\bf Theorem:} {\it Let $\phi: V_1 \to V_2$ be a non-trivial sheaf homomorphism between two semistable vector bundles with $\textnormal{rk} V_1 = \textnormal{rk} V_2$ and  $c_1(V_1)=c_1(V_2)$. If at least one of the bundles is stable, then $\phi$ is an isomorphism.}

\vspace{0.2cm}

When used with an appropriate correlation of the maps defining the two monads, this result is enough to show that the two bundles are the same. More precisely, there is a degree $(1,0)$ map in (\ref{VR2}) that we will, by using the freedom to choose coordinates, set equal to $x_0$ without loss of generality. With this choice made, the coefficients of $x_1$ in the two degree $(1,4)$ maps defining $V_D$ are set equal to the the coefficients of $x_1^2$ defining the two degree $(2,4)$ maps defining $V_R$ in (\ref{VR2}). Similarly, the coefficients of $x_1^2$ in the four degree $(2,3)$ maps appearing in the definition of $V_D$ are set equal to the coefficients of $x_1^3$ in the four degree $(3,3)$ maps defining $V_R$. With these correlations enforced, the two bundles are isomorphic. The two dual compactifications are, in this special case, in fact identical.

\vspace{0.2cm}

In this example, then, we have a pair of dual theories of the type we have been discussing which exhibit all of the same structure as the other cases of interest. In blindly trying to construct a mapping in field space such that the Yukawa couplings can be shown to be the same, one encounters all of the same difficulties as in a generic case of this type. In particular, this example is not of the special form that was considered in the last section. Nevertheless, in this case the Yukawa couplings {\it do} agree - simply because they are secretly the same example!

\vspace{0.3cm}

As a final comment for this section, we would like to observe that some of the above results are somewhat surprising. If two theories are dual, their low energy effective descriptions tend to agree in fairly non-trivial fashions. For example, in mirror symmetry, two $4D$ $N=2$  theories agree and the two dual descriptions cover the same portion of moduli space, but with a highly non-trivial field map between them. In the example of a flop in an $N=2$ type II string theory compactification, the two dual theories, related naively by a geometric transition, describe different parts of moduli space in their geometric regime. In addition, the matching of the two theories is complex. Non-perturbative effects resum in a non-trivial fashion to match perturbative ones on the other side of the duality. In the above, the matching of the two theories naively seems much more trivial. Perturbative effects are matching to perturbative effects, at least in the examples studied, and the mapping itself seems rather well behaved. The reason for this could well be that the quantities computed in this section do not depend on the nodal deformation of the quintic/the extra K\"ahler modulus on the resolution side of the duality. It is these fields which have an inverse relationship, tending to map regions in field space where supergravity is valid to regions where it is not. For example, a large K\"ahler modulus could map to a small coefficient for the nodal term in the quintic, which is close to a singularity of the variety as discussed earlier.

\section{Conclusions and future work}\label{concsec}
In this work we have taken the first steps to understanding whether recently discovered correspondences between 4-dimensional, ${\cal N}=1$ theories linked by Calabi-Yau geometric transitions are in fact new string dualities. While this question remains open, we have provided a partial mapping of 4-dimensional fields (based upon the geometry of the underlying manifolds, branes and bundles) and strong evidence for duality in the form of matching the holomorphic structure of the 4-dimensional, effective theory. For instance, in one illustrative example, we have shown that 147,440 independent superpotential Yukawa couplings match across the duality as functions of the moduli fields. 

There are a number of areas that it would be interesting to investigate in the future. The first of these is to ask whether we can complete the moduli mapping above to fully include the K\"ahler sector of the theory? This is largely unprobed by the current work because the perturbative superpotential effects we have studied in detail here do not depend on these moduli. One useful approach to this question would, therefore, be to consider non-perturbative corrections to the theory in the form of worldsheet instanton contributions. It is known that in the weakly coupled heterotic theory, worldsheet instantons can be generated by strings wrapping isolated genus zero curves in the Calabi-Yau manifold. The form of the non-perturbative superpotential contribution generated by strings wrapping such curves in a class $[C]$ is as follows.
\begin{equation} \label{mrnonpert}
W(C)=\sum_{{\cal C}_i \in [C]} f_i \, e^{-\alpha_a(C)T^a}
\end{equation}
Here the $f_i$'s are Pfaffian pre-factors which are functions of the bundle and complex structure moduli \cite{Braun:2008sf,Braun:2007vy,Braun:2007xh,Braun:2007tp}. The exponent is a complexification of the curve volume, expressed in terms of the K\"ahler moduli. Recent progress has made it computationally tractable to extract detailed information about the functional forms of the $f_i$ and to find large classes of examples in which (\ref{mrnonpert}) does not simply sum to zero \cite{Bertolini:2014dna,Buchbinder:2016rmw,Buchbinder:2017azb,Buchbinder:2018hns,Buchbinder:2019eal,Buchbinder:2019hyb} (avoiding the vanishing conditions of \cite{Beasley:2003fx,Distler:1992gi,Distler:1993mk,Silverstein:1995re,Basu:2003bq}).  It is clear that a comparison of such terms could lead to further information and constraints on the parts of the field mapping involving the K\"ahler sectors of the theories. Indeed, using the relationship between ``spectator" K\"ahler moduli given in \eref{boatymcequationface}, $T_D^a = T_R^a+f^a({T_R}^0)$ one could rewrite (\ref{mrnonpert}) for classes of curves $C$ which ``spectate" through the conifold transition and exist on both the deformation and resolutions manifolds. On the resolution side of the transition one could rewrite 
\begin{equation}
\sum_{{\cal C}_i \in [C]} {f^R}_i e^{-{\alpha^R}_a(C)T_R^a}~~ \Rightarrow ~~\sum_{{\cal C}_i \in [C]} {f^R}_i e^{-{\alpha^R}_a(C)(T_D^a-f^a(T_R^0))}
\end{equation}
Comparing the sum on the right hand side above, to contributions of the form $\sum_{{\cal C}_i \in [C]} {f^D}_{i}e^{-\alpha^D_a(C)T_D^a}$ on the deformation geometry could, when combined with the results on moduli mapping from this paper, specify key information about the free function $f^a({T_R}^0)$. We will undertake this comparison in future work \cite{nonpertpaper}. Note that such an approach assumes that perturbative and non-perturbative effects do not mix across the duality. While this is not usually the case in such correspondences (see for example Section 3.3 of \cite{Gendler:2022ztv}), the matching of perturbative results in this paper already hints that this might be the case here. Indeed, early results along these lines seem to suggest that at least some such matching of non-perturbative superpotential contributions does indeed occur across the duality being discussed in this paper \cite{nonpertpaper}.

Another important area of investigation would be to explore whether the correspondence we study here can be extended to other types of ${\cal N}=1$, 4-dimensional string compactifications. As pointed out in Section \ref{sec:Intro}, the structure of bridging branes described in Section \ref{sec:bridge_branes} is intrinsically tied to the geometry of conifold transitions themselves and not heterotic string theory in particular. It would would be interesting to investigate the structure of ${\cal N}=1$ effective theories arising from other types of string compactifications including those with spacetime filling branes available to wrap the bridging curves. These include Type I theories and IIB orientifolds with D5/O5 systems and F-theory on Calabi-Yau 4-folds (where the bridging objects analogous to the curves in the Calabi-Yau 3-fold case in Section \ref{sec:bridge_branes} are complex surfaces). We have begun preliminary explorations of such theories and their properties \cite{othertheories}.

Finally, perhaps the most important open question of this program is the following: If these theories are actually dual, what is the physical nature of the underlying duality? As discussed in Section \ref{sec:Intro}, one possibility is that the shared locus in geometry corresponding to the singular manifold (i.e. the conifold locus)  is not in fact a singular point in the field space of the 4-dimensional, ${\cal N}=1$ theory. That is, the additional components of the field space metric provided by the brane/bundle degrees of freedom could prevent the total field space metric from degenerating even though the Calabi-Yau geometry becomes singular. In this sense, the chains of manifold and bundle pairs connected by conifolds would all correspond to the same effective theory. This would be analgous to the correspondence between flop phases of Type IIB, ${\cal N}=2$ theories mentioned in Section \ref{sec:Intro}. Alternatively, analogous to Calabi-Yau Mirror symmetry, it would be remarkable if there existed a non-trival worldsheet automorphism of the underlying $(0,2)$ NLSMs which explained the correspondence we have noted here. In either case, to fully determine if such correspondences exist, it is necessary to have better computational control of the 4-dimensional theories near the conifold limit. We hope to return to such questions in future work.

\section*{Acknowledgements}

We would like to thank Callum Brodie for discussions and collaboration on the early stages of this project. All four authors are supported, in part, by NSF grant PHY-2310588.

\end{document}